\documentclass{aa}
\usepackage{psfig}
\def\TMB {$T_{\rm mb}$}

\def\TASTAR {$T_{\rm A}^{*}$}

\def\H0 {$H_{\rm o}$}

\def\SHUN {\hbox{$S_{100\mu \rm m}$}}
\def\solmass {\hbox{M$_{\odot}$}}
\def\solum {\hbox{L$_{\odot}$}}
\def\irlum {\hbox{$L_{\rm FIR}$}}

\def\numd {\hbox{$n({\rm H}_2$)}}                   

\def\kms {\hbox{${\rm km\,s}^{-1}$}}

\def\Kkms {\hbox{${\rm K\,km\,s}^{-1}$}}
\def\percc {$\hbox{{\rm cm}}^{-3}$}    
\def\cmsq  {$\hbox{{\rm cm}}^{-2}$}    

\def\ffas {\hbox{$\,.\!\!^{\prime\prime}$}}

\def\WAT {\hbox{${\rm H}_2{\rm O}$}}              
\def\CH3C2H {\hbox{${\rm CH}_3{\rm C}_2{\rm H}$}} 
\def\greekg1 {(\zeta _{\rm i},\eta _{\rm j})}

\begin{document}

\thesaurus{ 03                         
           (11.01.2                    
            11.09.1 M\,82              
            11.09.4                    
            11.14.1                    
            11.19.3                    
            13.19.1)}                  

\title{Dense gas in nearby galaxies
       \thanks {Based on observations with the Heinrich-Hertz-Telescope
               (HHT) and the IRAM 30-m telescope. The HHT is operated by 
               the Submillimeter Telescope Observatory on behalf of Steward 
               Observatory and the Max-Planck-Institut f{\"u}r Radioastronomie
               } }

\subtitle{XIII. CO submillimeter line emission from the starburst galaxy M\,82 }

\author{R.Q.~Mao\inst{1,2,3}, C.~Henkel\inst{1}, A.~Schulz\inst{4,5},
M.~Zielinsky\inst{6}, R.~Mauersberger\inst{7,8,9}, H.~St{\"o}rzer\inst{6},
T.L.~Wilson\inst{1,7}, \and P.~Gensheimer\inst{7}}

\offprints{C. Henkel; p220hen@mpifr-bonn.mpg.de}

\institute{
  Max-Planck-Institut f{\"u}r Radioastronomie,
  Auf dem H{\"u}gel 69, D-53121 Bonn, Germany
\and
  Purple Mountain Observatory, Chinese Academy of Sciences, 
  210008 Nanjing, PR China 
\and
  National Astronomical Observatories, Chinese Academy of Sciences,
  Beijing 100012, PR China
\and
  Institut f{\"u}r Physik und Didaktik, Universit{\"a}t zu K{\"o}ln, 
  Gronewaldstr. 22, D-50931 K{\"o}ln, Germany
\and
  Institut f{\"u}r Astrophysik und Extraterrestrische Forschung
  der Universit{\"a}t Bonn, Auf dem H{\"u}gel 71, D-53121 Bonn, Germany
\and
  I. Physikalisches Institut der Universit{\"a}t zu K{\"o}ln, Z{\"u}lpicher
  Stra{\ss}e 77, D-50937 K{\"o}ln, Germany
\and
  Submillimeter Telescope Observatory, The University of Arizona, Tucson
  AZ 85721, U.S.A.
\and
  Steward Observatory, The University of Arizona, Tucson AZ 85721, U.S.A.
\and
  Instituto de Radioastronomia Milim{\'e}trica, Avda. Divina Pastora, 7NC,
  E-18012 Granada, Spain
}

\titlerunning{Nucleus of M\,82}

\authorrunning{Mao et al.}

\date{Received date / Accepted date}

\maketitle

\markboth{R.Q. Mao, C. Henkel, A. Schulz et al.: 
          The nuclear region of the starburst galaxy M\,82}{}

\begin{abstract}

$^{12}$CO $J$ = 1--0, 2--1, 4--3, 7--6, and $^{13}$CO 1--0, 2--1, and 3--2 line
emission was mapped with angular resolutions of 13$''$ -- 22$''$ toward the
nuclear region of the archetypical starburst galaxy M\,82. There are two
hotspots on either side of the dynamical center, with the south-western lobe
being slightly more prominent. Lobe spacings are not identical for all
transitions: For the submillimeter CO lines, the spacing is $\sim$15$''$; for
the millimeter lines (CO $J$ = 2--1 and 1--0) the spacing is $\sim$26$''$,
indicating the presence of a `low' and a `high' CO excitation component.

A Large Velocity Gradient (LVG) excitation analysis of the submillimeter lines
leads to inconsistencies, since area and volume filling factors are almost
the same, resulting in cloud sizes along the lines-of-sight that match the 
entire size of the M\,82 starburst region. Nevertheless, LVG column densities
agree with estimates derived from the dust emission in the far infrared and
at submillimeter wavelengths. 22$''$ beam averaged total column densities 
are $N$(CO) $\sim$ 5\,10$^{18}$ and $N$(H$_2$) $\sim$ 10$^{23}$\,\cmsq; 
the total molecular mass is a few 10$^{8}$\,\solmass. 

Accounting for high UV fluxes and variations in kinetic temperature and
assuming that the observed emission arises from photon dominated regions 
(PDRs) resolves the problems related to an LVG treatment of the radiative 
transfer. Spatial densities are as in the LVG case (\numd\ $\sim$ 
10$^{3.7}$\,\percc\ and $\sim$10$^{3}$\percc\ for the high and low excitation 
component, respectively), but $^{12}$CO/$^{13}$CO intensity ratios 
$\ga$10 indicate that the bulk of the CO emission arises in UV-illuminated 
diffuse cloud fragments of small column density ($N$(H$_2)$ $\sim$ 
5\,10$^{20}$\,\cmsq/\kms) and sub-parsec cloud sizes with area filling factors 
$\gg$1. Thus CO arises from quite a different gas component than the classical
high density tracers (e.g. CS, HCN) that trace star formation rates more 
accurately. The dominance of such a diffuse molecular interclump medium 
also explains observed high [C\,{\sc i}]/CO line intensity ratios. PDR 
models do not allow a determination of the relative abundances of $^{12}$CO 
to $^{13}$CO. Ignoring magnetic fields, the CO emitting gas appears to be 
close to the density limit for tidal disruption. Neither changes in the 
$^{12}$C/$^{13}$C abundance ratio nor variations of the incident far-UV 
flux provide good fits to the data for simulations of larger clouds.

A warm diffuse ISM not only dominates the CO emission in the starburst
region of M\,82 but is also ubiquitous in the central region of our Galaxy,
where tidal stress, cloud-cloud collisions, shocks, high gas pressure, and
high stellar densities may all contribute to the formation of a highly
fragmented molecular debris. $^{12}$CO, $^{12}$CO/$^{13}$CO, and [C\,{\sc
i}]/CO line intensity ratios in NGC\,253 (and NGC\,4945) suggest that the CO
emission from the centers of these galaxies arises in a physical environment
that is similar to that in M\,82. Starburst galaxies at large distances ($z$
$\sim$ 2.2--4.7) show $^{12}$CO line intensity ratios that are consistent with
those observed in M\,82. PDR models should be applicable to all these sources.
$^{12}$CO/$^{13}$CO line intensity ratios $\gg$10, sometimes observed in
nearby ultraluminous mergers, require the presence of a particularly diffuse,
extended molecular medium. Here [C\,{\sc i}]/CO abundance ratios should be
as large or even larger than in M\,82 and NGC\,253.

\ \ \ \ \\

\keywords{
   Galaxies: active -- Galaxies: ISM -- Galaxies: nuclei --
   Galaxies: starburst -- Galaxies: individual (M\,82) -- 
   Radio lines: galaxies
}

\end{abstract}

\section{Introduction} 

Low lying rotational transitions of CO are widely used as tracers of molecular
hydrogen and are essential to determine dynamical properties and total
molecular masses of galaxies. The widespread use of CO $J$ = 1--0 and 2--1
spectroscopy is however not sufficiently complemented by systematic surveys in
higher rotational CO transitions to confine the excitation conditions of the
dense interstellar medium (ISM). While the $J$ = 1 and 2 states of CO are only
5.5 and 17\,K above the ground level, the $J$ = 3 to 7 states are at 33, 55,
83, 116, and 155\,K and trace a component of higher excitation. `Critical
densities', at which collisional deexcitation matches spontaneous decay in the
optically thin limit, are $\sim$ 10$^{5-6}$\,\percc\ for CO $J$ = 3--2 to 7--6
in contrast to 10$^{3.5}$ and 10$^{4.3}$\,\percc\ for the ground rotational CO
transitions.

Starburst galaxies are known to contain large amounts of molecular gas that
may be heated to $T_{\rm kin}$ $\sim$ 100\,K by young massive stars, cosmic
rays or turbulent heating. Therefore highly excited CO transitions, observed at
submm-wavelengths, are {\it the} appropriate tool to study this interstellar
gas component. Among the three nearest ($D$ $\sim$ 3\,Mpc) nuclear
starburst galaxies, NGC\,253, NGC\,4945, and M\,82 (NGC\,3034) M\,82 is most
readily accessible from telescopes of the northern hemisphere. Containing one
of the brightest IRAS point sources beyond the Magellanic Clouds
(\SHUN\,\,$\sim$ 1000\,Jy), M\,82 has been observed at a variety of wavelengths,
ranging from the radio to the $\gamma$-ray domain of the electromagnetic
spectrum. The starburst in M\,82 is likely triggered by a tidal interaction
with M\,81, causing a high infrared luminosity (\irlum\,\,$\sim$
4\,10$^{10}$\,\solum), a high density of supernova remnants, and copious
amounts of dense gas with strong OH and \WAT\ masers and a large number of
molecular high density tracers (for CO maps, see Sutton et al. 1983; Olofsson
\& Rydbeck 1984; Young \& Scoville 1984; Nakai et al. 1986, 1987; Lo et al.
1987; Loiseau et al. 1988, 1990; Phillips \& Mampaso 1989; Turner et al. 1991;
Tilanus et al. 1991; Sofue et al. 1992; White et al. 1994; Shen \& Lo 1995;
Kikumoto et al. 1998; Neininger et al. 1998).

So far, few CO 4--3 maps of external galaxies were published (for M\,51,
M\,82, M\,83, and NGC6946 see White et al. 1994; Petitpas \& Wilson 1998;
Nieten et al. 1999). Among these M\,82 is the only true starburst galaxy but
its CO 4--3 map (White et al. 1994) is confined to the very central region.
With respect to higher rotational CO transitions, only a few CO 6--5 spectra
were presented from nearby galaxies (Harris et al. 1991; Wild et al. 1992).

We have used the Heinrich-Hertz-Telescope (HHT) on Mt. Graham (Baars \& Martin
1996) to map M\,82 in the CO $J$= 7--6, 4--3, and $^{13}$CO 3--2 transitions.
These data are complemented by new $J$ = 2--1 and 1--0 spectra taken with the
IRAM 30-m telescope.

\section{Observations} 

\subsection{Observations with the Heinrich-Hertz-Telescope}

$^{13}$CO 3--2 (331\,GHz $\simeq$ 907$\mu$m), $^{12}$CO 4--3 (461\,GHz
$\simeq$ 650$\mu$m), and $^{12}$CO 7--6 (807\,GHz $\simeq$ 372$\mu$m) line
emission was observed at the HHT during Feb. 1999 with beamwidths of $\sim$
22$''$, 18$''$, and 13$''$, respectively. For the CO 3--2 and 4--3
transitions, SIS receivers were employed; the CO 7--6 line was observed with a
Hot Electron Bolometer (HEB) kindly provided by the Center for Astrophysics
(Kawamura et al. 1999). The backends consisted of two acousto optical
spectrometers, each with 2048 channels (channel spacing $\sim$ 480\,kHz,
frequency resolution $\sim$ 930\,kHz) and a total bandwidth $\sim$ 1\,GHz.

Spectra were taken using a wobbling (2\,Hz) secondary mirror with a beam
throw of $\pm$120 to $\pm$240$''$ in azimuth. Scans obtained with reference
positions on either azimuth were coadded to ensure flat baselines. Receiver
temperatures were at the order of 170\,K at 331\,GHz, 150\,K at 461\,GHz, and
1000\,K at 807\,GHz; system temperatures were $\sim$ 900, 3500, and 8000\,K on
a \TASTAR\ scale, respectively.

The receivers were sensitive to both sidebands. Any imbalance in the gains of
the lower and upper sideband would thus lead to calibration errors. To account
for this, we have observed Orion-KL (at 807\,GHz) and IRC+10216 (at 807 and
331\,GHz) prior to our M\,82 measurements with the same receiver tuning setup.
Peak temperatures were 70, 20, and 2.3\,K on a \TASTAR\ scale, respectively
(cf. Howe et al. 1993; Groesbeck et al. 1994). At 461\,GHz, Orion-KL (\TASTAR\
$\sim$ 70\,K; cf. Schulz et al. 1995) was mapped but the tuning was later
changed by $\Delta V$ = 200\,\kms\ for M\,82 (see Sect.\,3.2).

All results displayed are given on a main beam brightness temperature scale
(\TMB). This is related to \TASTAR\ via \TMB\,\,= \TASTAR\ ($F_{\rm
eff}$/$B_{\rm eff}$) (see Downes 1989). Main beam efficiencies, $B_{\rm eff}$,
were 0.5, 0.38, and 0.36 at 330, 461, and 806\,GHz, as obtained from
measurements of Saturn. Forward hemisphere efficiencies are 0.9, 0.75, and
0.70, respectively (D. Muders, priv. comm.). With an rms surface accuracy 
of $\sim$ 20$\mu$m ($\lambda$/18 at 806\,GHz), the HHT is quite accurate. This
reduces the effect of the source coupling efficiency on the measured source
size.

At the center of M\,82, CO lineshapes depend sensitively on the position
observed, so that in each of our maps the dynamical center could be identified
with an accuracy better than 5$''$. While relative pointing errors should
be small when compared to the spacing (10$''$) of our $^{13}$CO $J$ =
3--2 and CO 4--3 maps, relative pointing is less reliable in the case of 
the CO $J$ = 7--6 map; here deviations may reach $\pm$5$''$ for a few
positions.

\begin{figure*}
\hspace{-0.3cm}
\psfig{figure=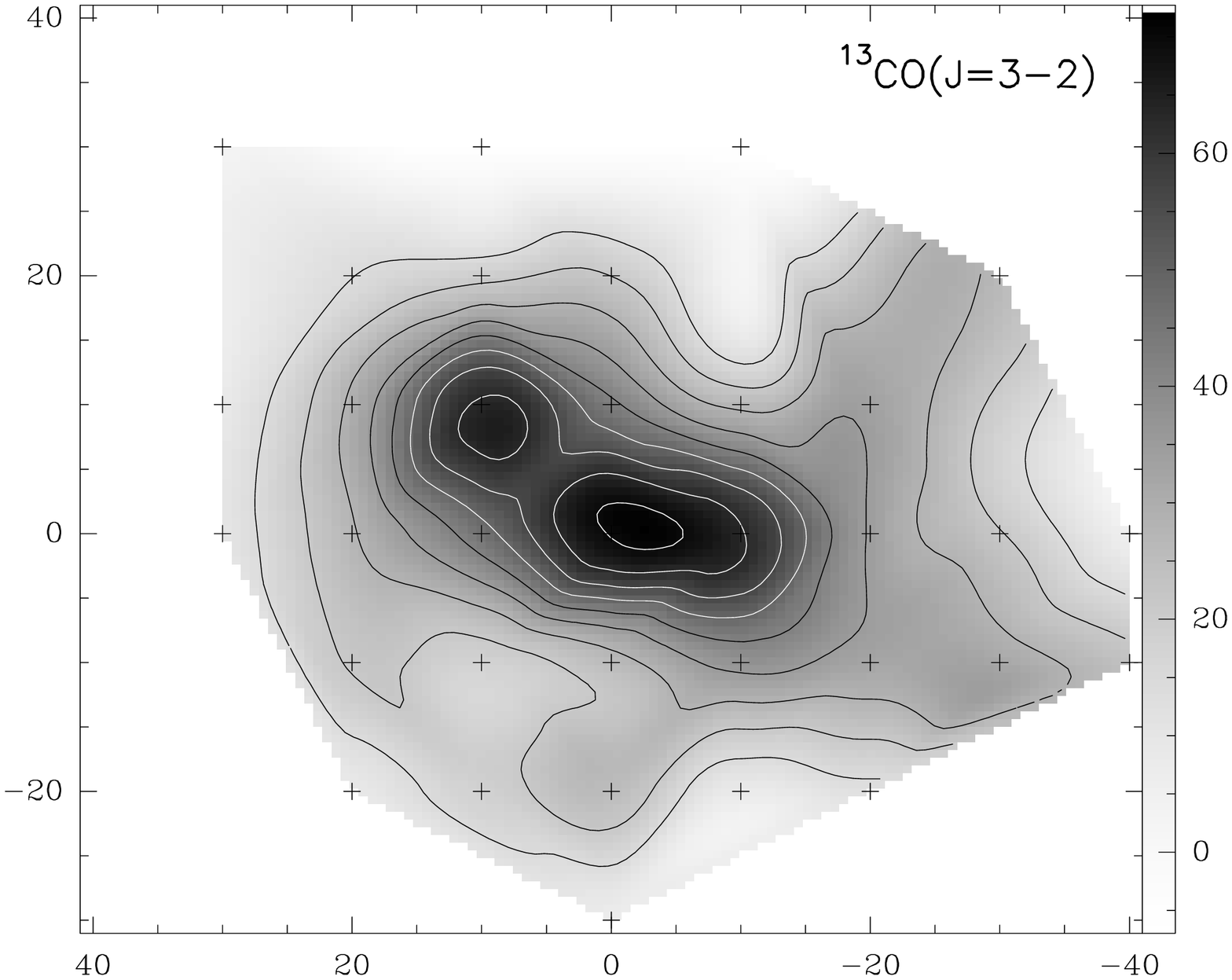,height=8.0cm,angle=-90}
\hspace{-2.5cm}
\psfig{figure=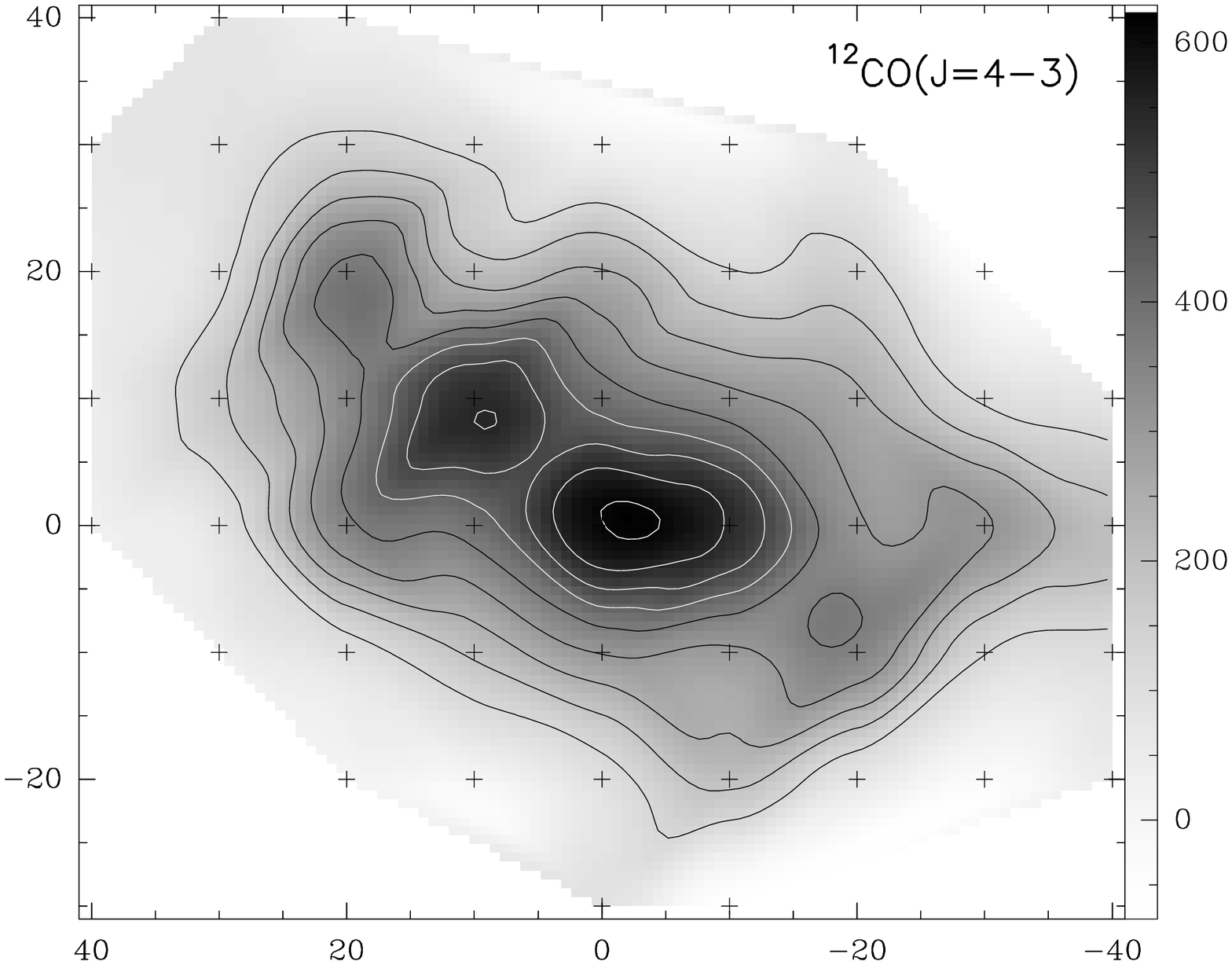,height=8.0cm,angle=-90}
\end{figure*}

\begin{figure*}
\hspace{-0.30cm}
\psfig{figure=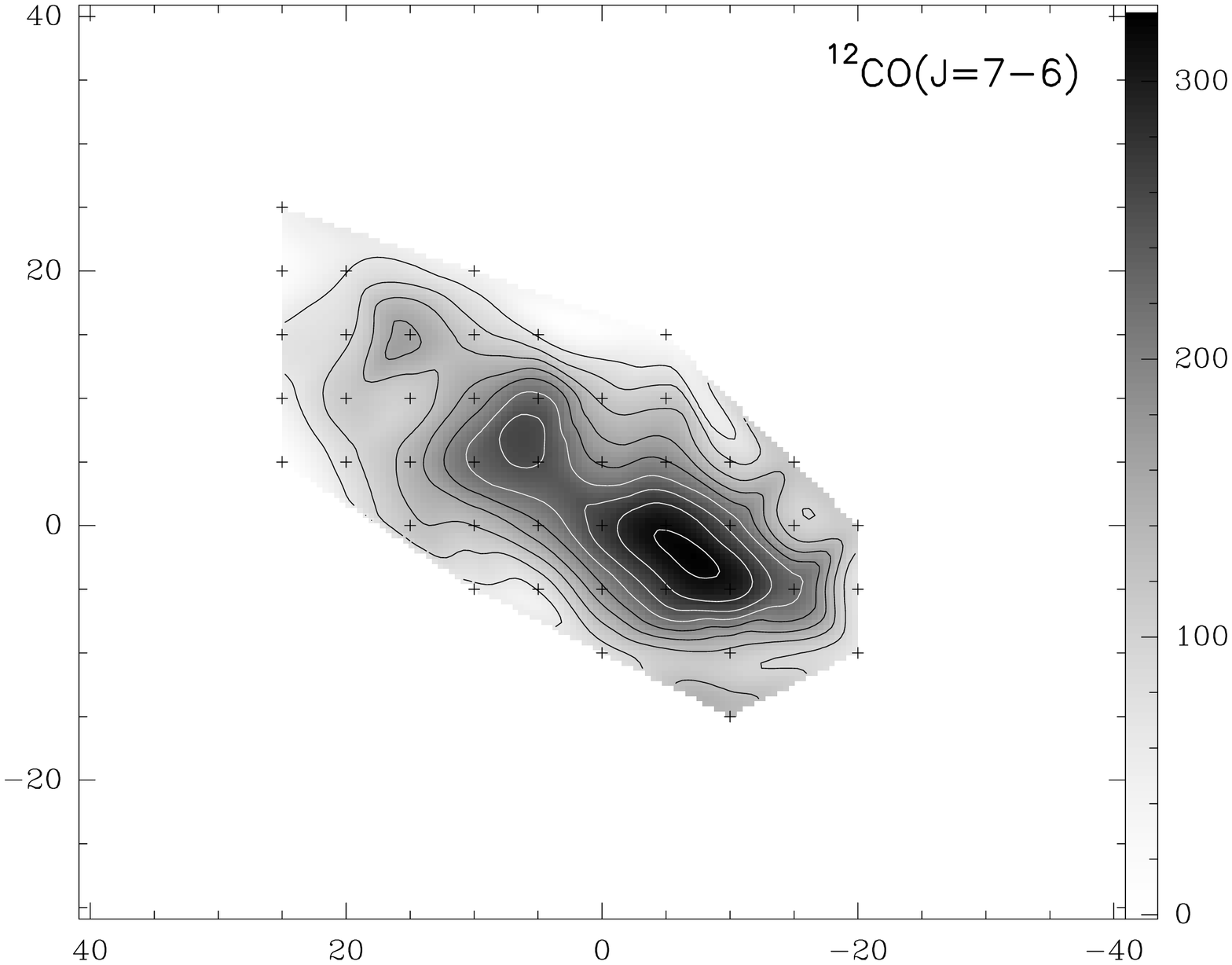,height=8.0cm,angle=-90}
\hspace{-1.8cm}
\begin{minipage}[c]{.45\textwidth}
 \centering
 \vspace{-8.5cm}
 \caption[]{
Intensities integrated over the velocity interval 25 to 385\,\kms\ towards
the nuclear starburst region of M\,82. 

{\it Upper left}: $^{13}$CO $J$ = 3--2 map over the central
50$''$ $\times$ 50$''$. Contour levels are 14.2 to 71.0 by 7.1\,\Kkms. 
 
{\it Upper right}: $^{12}$CO $J$ = 4--3 map over the central
80$''$ $\times$ 70$''$. Contour levels are 124 to 620 by 62\,\Kkms.

{\it Lower left}: $^{12}$CO $J$ = 7--6 map over the central
40$''$ $\times$ 30$''$. Contour levels are 64 to 320 by 32\,\Kkms. 

The nominal reference position is $\alpha_{1950}$ =
9$^{\rm h}$ 51$^{\rm m}$ 43$^{\rm s}$, $\delta_{1950}$ = 69$^{\circ}$
55$'$ 00$''$, but in view of pointing errors (see Sects.\,2.1 and 4.1),
coordinates are best defined by a comparison with interferometric maps (e.g.
Lo et al. 1987; Shen \& Lo 1995; Neininger et al. 1998). For consistency with
other studies, $D$ = 3.25\,Mpc is assumed throughout the paper. Recent distance
determinations indicate slightly higher values (e.g. $D$ = 3.9$\pm$0.3\,Mpc;
Sakai \& Madore 1999).
}
\label{m82-co7-6-tmb-contour-grey}
\end{minipage}
\end{figure*}

\subsection{Observations with the IRAM 30-m telescope}

$^{12}$CO and $^{13}$CO $J$ = 1--0 and 2--1 observations were made with SIS
receivers of high image sideband rejection ($\sim$25\,db for 1--0 and
$\sim$13\,db for 2--1 line data) of the inner 50$''$ $\times$ 50$''$
($^{12}$CO; 5$''$ spacing for the central 20$''$, otherwise 10$''$) and
20$''$ $\times$ 20$''$ ($^{13}$CO; 5$''$ spacing) of M\,82 in June 1999. The
measurements were made in a position switching mode with the off-position
displaced by 15$'$ in right ascension. The two $^{12}$CO lines (as well as
those of $^{13}$CO) were measured simultaneously. Beamwidths at 115 ($J$ =
1--0) and 230\,GHz ($J$ = 2--1) were 21 and 13$''$; forward hemisphere
and beam efficiencies were 0.92 and 0.72 for the $J$ = 1--0 and 0.89 and
0.45 for the $J$ = 2--1 data, respectively. Calibration was checked by
observing IRC+10216. Measured line intensities were $T_{\rm mb}$ = 17 and
43\,K for $^{12}$CO $J$ = 1--0 and 2--1 (channel spacings were 3.2 and
1.6\,\kms, respectively) and 2.3 and 6.0\,K for $^{13}$CO $J$ = 1--0 and 2--1
(channel spacings: 2.7 and 1.4\,\kms; cf. Mauersberger et al. 1989).

\begin{figure}
\hspace{-0.8cm}
\psfig{figure=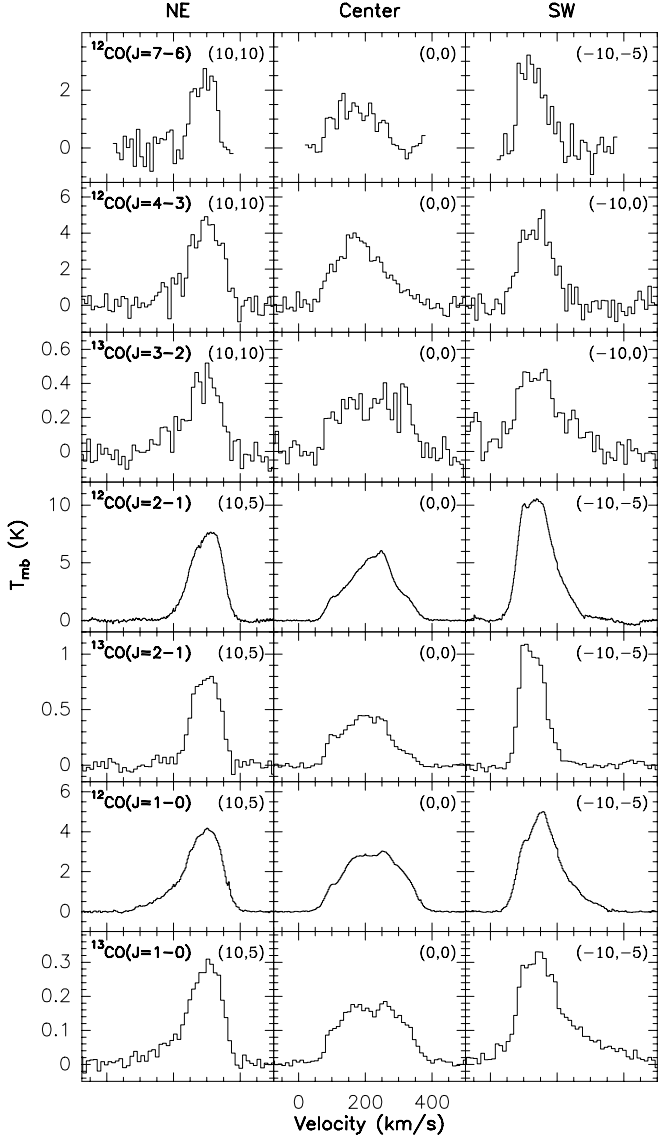,height=17.0cm,angle=0}
\vspace{-0.5cm} 
\caption[]{Spectra towards the center (middle), the north-eastern (left), and
the south-western (right) lobe. The temperature is given in units of main beam
brightness temperature (K). Beamwidths are 13$''$, 18$''$, 22$''$, 13$''$,
13$''$, 21$''$ and 21$''$ for the $^{12}$CO 7--6, $^{12}$CO 4--3, $^{13}$CO
3--2, $^{12}$CO and $^{13}$CO 2--1, and $^{12}$CO and $^{13}$CO 1--0 profiles,
respectively. Offsets in arcsec relative to a nominal center position (see 
Fig.\,\ref{m82-co7-6-tmb-contour-grey}) are given in the upper right corner
of each box. Small differences in these offsets are caused by pointing
deviations (Sect.\,2) or changes in source morphology (Sect.\,4). All data
are smoothed to a channel width of $\sim$10\,\kms. Linewidths (compare e.g.
the $J$ = 2--1 and 7--6 profiles from the south-western lobe with the
corresponding $J$ = 3--2 and 1--0 spectra) are affected by differences in
beamsize.
}
\label{m82-three-comp-all} 
\end{figure}

\begin{figure}
\vspace{0.0cm}
\hspace{0.3cm}
\psfig{figure=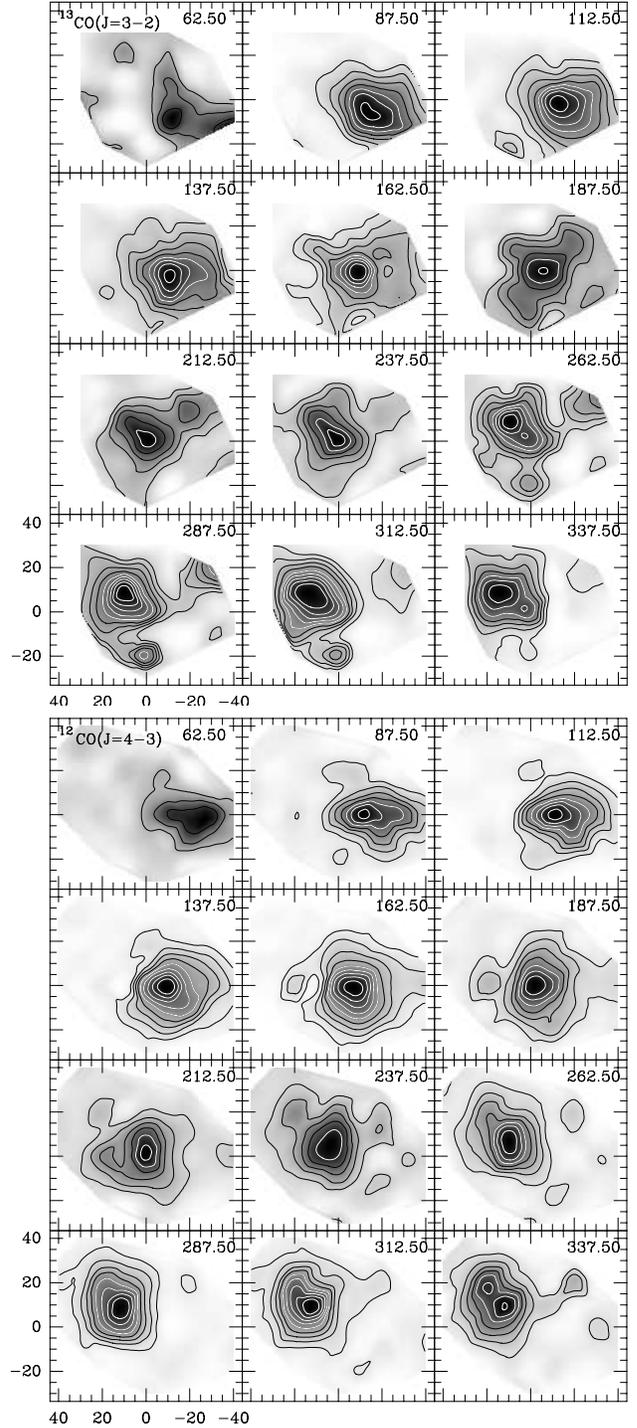,height=19.0cm}
\vspace{0.0cm}
\caption[]{$^{13}$CO $J$ = 3--2 ({\it upper panel}) and $^{12}$CO $J$ = 4--3
({\it lower panel}) channel maps. Center velocities are given in the upper 
right corner of each box. The radial velocity range is 30\,\kms\ per image.
The lowest black contour, the lowest white contour and the interval are 
respectively 1.4\,\Kkms, 7\,\Kkms, and 1.4\,\Kkms\ ({\it upper panel}) and 
14\,\Kkms, 70\,\Kkms, and 14\,\Kkms\ {(\it lower panel}) on a $T_{\rm mb}$
scale. 
} 
\label{m82-co-ch-map-grey}
\end{figure}

\begin{figure}
\vspace{-1cm}
\hspace{-0.1cm}
\psfig{figure=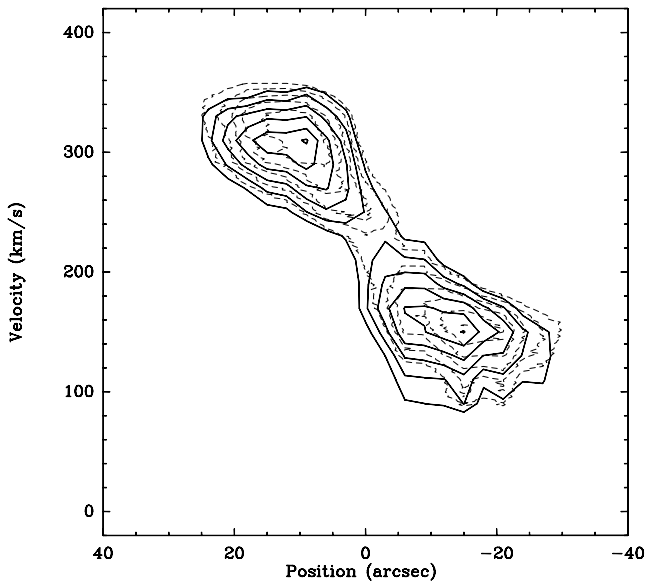,height=9.0cm,angle=0}
\caption[]{Position-velocity diagram for CO $J$ = 4--3 (solid lines; our data) 
and 2--1 (dashed lines; data from Wei{\ss} (in preparation)) along the line
connecting the south-western and north-eastern lobe. Angular resolutions are 
18$''$ and 13$''$, respectively. The zero offset position is located half way 
between the emission peaks. CO contour levels are 50\%, 60\%, ..., 90\%, and 
99\% of the respective peak intensities (see Table\ \ref{tab:intensities}).}
\label{m82-pv}
\end{figure}

\section{Results}

\subsection{Overall morphology of CO submillimeter line emission}

Maps of integrated $^{13}$CO $J$ = 3--2, $^{12}$CO 4--3, and $^{12}$CO 7--6
intensity are presented in Fig.\,\ref{m82-co7-6-tmb-contour-grey}. Detectable 
emission is strongly confined to the central part of the galaxy. In each map 
a particularly wide spectral feature, with slightly smaller peak intensity 
than the most intense lines in the SW and NE, could be identified with the 
dynamical center (see Fig.\,\ref{m82-three-comp-all} which also contains 
IRAM spectra). Two main peaks of emission are detected, being displaced by 
almost 10$''$ from the center (this corresponds to a projected distance of 
150\,pc); the south-western hot spot is most prominent, while evidence for a 
third peak in the NE (at $\sim$ (17$''$,17$''$)), seen in the CO $J$ = 4--3 
and 7--6 maps, is not conclusive. Velocity channel maps of the $^{13}$CO 
$J$ = 3--2 and $^{12}$CO $J$ = 4--3 line emission are displayed in
Fig.\,\ref{m82-co-ch-map-grey}. These outline the extent of the emission 
at various velocities, show the dominant rotation pattern with the red-shifted 
lobe in the north-east and the blue-shifted lobe in the south-west, and 
indicate the rapid change in radial velocity near the dynamical center 
(cf. Neininger et al. 1998). 

\subsection{Consistency of CO line temperatures} 

Calibration at sub-millimeter wavelengths is critical because of rapidly
changing weather conditions, high atmospheric opacities, imbalances in the
receiver gains between the sidebands, and because of uncertainties in beam and
forward hemisphere efficiencies. Calibration uncertainties introduced by these
effects may rise up to $\pm$30\% and a comparison with data published
elsewhere is needed. Intensities of our $^{13}$CO $J$ = 3--2 spectra from the
center, the south-western, and north-eastern lobes are smaller by $\sim$30\%
than those given by Tilanus et al. (1991; their Fig.\,3). Our spectrum from
IRC+10216 is however 30\% stronger (on a $T_{\rm mb}$ scale) than that given
by Groesbeck et al. (1994), so that our scaling is intermediate between those
of Tilanus et al. and Groesbeck et al. CO $J$ = 4--3 line intensities
(Fig.\,\ref{m82-three-comp-all}) are $\sim$30\% larger than those
given by White et al. (1994). Since their data were obtained with higher
angular resolution (JCMT beamwidth: 11$''$), this difference is significant. A
comparison with the three 4--3 spectra from the 10-m CSO shown by G{\"u}sten
et al. (1993) shows good agreement for the lobe positions. Differences in 4--3
peak temperatures between central and lobe positions are however less
pronounced than reported by G{\"u}sten et al. (we find peak line temperature
ratios of $\sim$1.3 instead of $\sim$2.0 between the lobes and the center).
Our CO 7--6 line intensities from Orion-KL agree well with those obtained by
Howe et al. (1993) with the 10-m CSO antenna.

Previous high angular resolution CO data from the 30-m telescope (compare
Fig.\,9a of Loiseau et al. 1990 with our Table\ \ref{tab:intensities}) were
affected by uncertainties in the image sideband ratios. Our new data are not
significantly affected by this problem, should be reliably calibrated within
$\pm$10\%, and agree well with recent spectra obtained independently in an
`on-the-fly' observing mode (Wei{\ss}, in preparation).

\begin{table*}
\caption[]{\label{tab:intensities} Line intensities ($T$ = $T_{\rm mb}$), 
integrated line intensities ($I$ = $\int{T_{\rm mb} {\rm d}v}$), and 
line temperature ratios for a beamwidth of 22$''$\,\,$^{a}$. CO $J$ = 1--0
and 2--1 data were obtained with the IRAM 30-m telescope, higher excited
rotational lines were measured with the HHT. }
\begin{flushleft}
\begin{tabular}{llccc}
\hline
\multicolumn{2}{c}{Line or Line Ratio} & 
\multicolumn{3}{c}{Position$^{b}$} \\
                            &           &     NE-Lobe    &     Center     &
                                              SW-Lobe   \\
\hline
                            &           &                &                &
                                                        \\
{\it Line temperatures}:    &           &                &                &
                                                        \\
$T$(CO $J$ = 7--6)          & (K)       & 1.08$\pm$0.11  & 0.82$\pm$0.11  & 
                                          1.37$\pm$0.11 \\ 
$T$(CO $J$ = 4--3)          & (K)       & 3.40$\pm$0.18  & 2.82$\pm$0.18  & 
                                          3.72$\pm$0.18 \\ 
$T$(CO $J$ = 2--1)          & (K)       & 4.16$\pm$0.03  & 4.26$\pm$0.04  & 
                                          5.33$\pm$0.03 \\ 
$T$(CO $J$ = 1--0)          & (K)       & 4.18$\pm$0.03  & 3.14$\pm$0.02  & 
                                          5.00$\pm$0.03 \\ 
$T$($^{13}$CO $J$ = 3--2)   & (K)       & 0.40$\pm$0.03  & 0.28$\pm$0.03  & 
                                          0.41$\pm$0.03  \\ 
$T$($^{13}$CO $J$ = 2--1)$^{c}$& (K)    & 0.46$\pm$0.02  & 0.31$\pm$0.02  & 
                                          0.58$\pm$0.02 \\ 
$T$($^{13}$CO $J$ = 1--0)   & (K)       & 0.30$\pm$0.02  & 0.19$\pm$0.01  & 
                                          0.30$\pm$0.02 \\ 
                            &           &                &                &
                                                        \\
{\it Integrated Intensities}:&          &                &                &
                                                         \\
$I$(CO $J$ = 7--6)          & (\Kkms)   & 132.3$\pm$5.8  & 177.7$\pm$5.4  & 
                                          167.2$\pm$6.4  \\
$I$(CO $J$ = 4--3)          & (\Kkms)   & 465.3$\pm$12.9 & 493.3$\pm$12.1 &
                                          503.0$\pm$12.6 \\
$I$(CO $J$ = 2--1)          & (\Kkms)   & 674.3$\pm$0.8  & 847.9$\pm$1.2  & 
                                          804.9$\pm$1.1  \\
$I$(CO $J$ = 1--0)          & (\Kkms)   & 573.9$\pm$0.6  & 663.0$\pm$0.6  & 
                                          672.8$\pm$0.8  \\
$I$($^{13}$CO $J$ = 3--2)   & (\Kkms)   &  51.2$\pm$4.9  &  71.7$\pm$3.8  &
                                           68.0$\pm$2.8  \\
$I$($^{13}$CO $J$ = 2--1)$^{c}$& (\Kkms)&  58.9$\pm$1.1  &  59.8$\pm$0.9  & 
                                           69.0$\pm$1.1  \\
$I$($^{13}$CO $J$ = 1--0)   & (\Kkms)   &  40.1$\pm$0.5  &  41.6$\pm$0.3  & 
                                           58.8$\pm$0.6  \\
                            &           &                &                &
                                                        \\
{\it Ratios}:               &           &                &                &
                                                         \\
$T$(CO $J$ = 7--6/CO $J$ = 4--3)&       & 0.32$\pm$0.03  & 0.29$\pm$0.04  &
                                          0.37$\pm$0.03  \\
$T$(CO $J$ = 4--3/$^{13}$CO $J$ = 3--2)&& 8.50$\pm$0.87  & 10.1$\pm$1.26  &
                                          9.07$\pm$0.80  \\
$T$(CO $J$ = 2--1/CO $J$ = 1--0)&       & 1.00$\pm$0.01  & 1.36$\pm$0.02  &
                                          1.07$\pm$0.01  \\
$T$($^{13}$CO $J$ = 3--2/$^{13}$CO $J$ = 2--1)$^{c}$     &
                                        & 0.87$\pm$0.08  & 0.90$\pm$0.11  &
                                          0.71$\pm$0.06  \\
$T$($^{13}$CO $J$ = 2--1/$^{13}$CO $J$ = 1--0)$^{c}$     &
                                        & 1.53$\pm$0.08  & 1.63$\pm$0.06  &
                                          1.93$\pm$0.04  \\
$T$(CO $J$ = 2--1/$^{13}$CO $J$ = 2--1)$^{c}$            &
                                        & 9.04$\pm$0.40  & 13.7$\pm$0.90  &
                                          9.19$\pm$0.32  \\
$T$(CO $J$ = 1--0/$^{13}$CO $J$ = 1--0)$^{c}$            &
                                        & 13.9$\pm$0.93  & 16.5$\pm$0.88  &
                                          16.7$\pm$1.12  \\
                            &           &                &                &
                                                         \\
$I$(CO $J$ = 7--6/CO $J$ = 4--3)&       & 0.28$\pm$0.01  & 0.36$\pm$0.01  &
                                          0.33$\pm$0.02  \\
$I$(CO $J$ = 4--3/$^{13}$CO $J$ = 3--2)&& 9.09$\pm$0.91  & 6.88$\pm$0.40   &
                                          7.40$\pm$0.36  \\
$I$(CO $J$ = 2--1/CO $J$ = 1--0)&       & 1.17$\pm$0.01  & 1.28$\pm$0.01  &
                                          1.20$\pm$0.01  \\
$I$($^{13}$CO $J$ = 3--2/$^{13}$CO $J$ = 2--1)$^{c}$     & 
                                        & 0.87$\pm$0.08  & 1.20$\pm$0.06  &
                                          0.99$\pm$0.04  \\
$I$($^{13}$CO $J$ = 2--1/$^{13}$CO $J$ = 1--0)$^{c}$     & 
                                        & 1.47$\pm$0.03  & 1.44$\pm$0.02  &
                                          1.17$\pm$0.02  \\
$I$(CO $J$ = 2--1/$^{13}$CO $J$ = 2--1)$^{c}$            &
                                        & 11.4$\pm$0.21  & 14.2$\pm$0.21  &
                                          11.7$\pm$0.19  \\
$I$(CO $J$ = 1--0/$^{13}$CO $J$ = 1--0)$^{c}$            &
                                        & 14.3$\pm$0.18  & 15.9$\pm$0.12  &
                                          11.4$\pm$0.12  \\
\hline
\end{tabular}
\end{flushleft}
a) Given 1$\sigma$ errors in $T_{\rm mb}$ and $\int{T_{\rm mb} 
   {\rm d}v}$ were determined from gaussian fits and neither 
   include calibration uncertainties (see Sects.\,2 and 3.2) nor 
   errors caused by the limited spatial coverage of our CO 7--6 
   (Fig.\,\ref{m82-co7-6-tmb-contour-grey}), $^{13}$CO 2--1, and 
   $^{13}$CO 1--0 (Sect.\,2) maps. For the beam convolution, gaussian 
   distributions were assumed for the telescope beam and the source. 
   While all measured positions were taken for beam
   convolution, the CO $J$ = 4--3 beam size is almost as large as that
   for the $^{13}$CO $J$ = 3--2 spectra. Therefore only five measured CO 
   $J$ = 4--3 spectra, the profile measured toward the specific position 
   and those with offsets of 10$''$ along the four cardinal directions, 
   dominate the convolution. For the $J$ = 7--6 line, spatially 
   constraining the convolution to positions with weights $>$0.3 
   or $\Delta$ = 11\ffas7 (weight = 
   exp[--4\,\,ln\,2\,($\Delta^{2}$/($\theta_{\rm b1}^{2} - 
   \theta_{b2}^{2})$)]; $\Delta$: Offset; $\theta_{\rm b1}$ = 22$''$, 
   $\theta_{\rm b2}$ = 18$''$ or 13$''$ for $J$ = 4--3 and 7--6 spectra, 
   respectively) modifies the result by only $\sim$5\%. \\
b) For the offsets in arcsec, see Fig.\ref{m82-three-comp-all}.\\
c) Because of the limited extent of the $^{13}$CO $J$ = 2--1 map (Sect.\,2.2),
   $^{13}$CO 2--1 values for the lobes should be considered with caution.
   The problem is less severe for the $^{13}$CO $J$ = 1--0 transition, 
   because beam convolution is not required in this case.
   \\
\end{table*}
 
\subsection{Line intensity ratios}

Claims that the large scale ($\ga$20$''$) integrated CO $J$ = 2--1/1--0 line
intensity ratio were much larger than unity (e.g. Knapp et al. 1980; Olofsson
\& Rydbeck 1984; Loiseau et al. 1990) can be firmly rejected. The more
recently measured ratios of 1.0 (Wild et al. 1992), 1.1 (Mauersberger et al.
1999) and 1.0--1.4 (Table\ \ref{tab:intensities}) imply that $^{12}$CO line
intensities of the three lowest rotational CO transitions must be similar over
the central 22$''$. Our $^{13}$CO $J$ = 3--2 line temperatures are a factor of
$\sim$10 smaller than those of the $J$ = 3--2 $^{12}$CO transition observed with
the same telescope (R. Wielebinski, priv. comm.). The situation with respect to 
the $J$ = 6--5 line remains unclear. 6--5 spectra from the western lobe 
(Harris et al. 1991) and the center (Wild et al. 1992; their Fig.\,2) do not 
allow a beam convolution to the angular resolution of the lower frequency data 
and the beam pattern appears to be complex. Our $J$ = 7--6 data show peak line 
temperatures of order 2--3\,K. This is smaller than the 4\,K measured in the 
$J$ = 4--3 and lower $J$ transitions with larger beam sizes. Beyond $J$ = 
4--3 we thus find {\it clear evidence for a weakening of CO emission with 
increasing rotational quantum number $J$}.

\section{Spatial distributions: Are there differences?}

\subsection{CO}

A comparison of the CO $J$ = 4--3 data presented in
Fig.\,\ref{m82-co7-6-tmb-contour-grey} with those of White et al. (1994) shows
a strong discrepancy in the overall spatial distribution: Our CO $J$ = 4--3
map contains (at least) two maxima of emission, while the higher angular
resolution data of White et al. (1994) have only one peak. Data with only one
peak (that are based on spectra with sufficient resolution to separate the
lobes) are also presented by Wild et al. (1992; their Fig.\,10 and Table 5)
for $^{13}$CO and C$^{18}$O $J$ =2--1. The original $^{13}$CO $J$ = 2--1
spectra displayed by Loiseau et al. (1988; their Fig.\,1), however, clearly
show a double-lobed distribution. Maps in the low-$J$ $^{12}$CO and $^{13}$CO
transitions (e.g. Lo et al. 1987; Nakai et al. 1987; Loiseau et al. 1988,
1990; Tilanus et al. 1991; Shen \& Lo 1995; Neininger et al. 1998) as well as
our $J$ = 7--6 (Fig.\,3), 2--1 and 1--0 data {\it all} show a double-lobed
structure. We thus conclude that, in spite of previous evidence to the
contrary, {\it the overall spatial distribution of emission from highly
excited CO shows two main centers of emission}.

\begin{table}
\caption[]{\label{lobes} Lobe separations$^{a)}$ }
\begin{flushleft}
\begin{tabular}{lrl}
\hline
                               &            &                            \\
Tracer                         & Separation & Reference                  \\
                               & arcsec     &                            \\
                               &            &                            \\
\hline
                               &            &                            \\
{\it Continuum}                &            &                            \\
2.2$\mu$m                      &  9$\pm$1   & McLeod et al. (1993)$^{b}$ \\
2.2$\mu$m                      &  8$\pm$2   & Dietz et al. (1986)        \\
10.0$\mu$m                     &  9$\pm$1   & Dietz et al. (1989)        \\
10.8$\mu$m                     &  9$\pm$1   & Telesco et al. (1991)      \\
12.4$\mu$m                     &  9$\pm$1   & Telesco \& Gezari (1992)   \\
19.2$\mu$m                     & 11$\pm$2   & Telesco et al. (1991)      \\
19.5$\mu$m                     & 10$\pm$2   & Dietz et al. (1989)        \\
30.0$\mu$m                     & 10$\pm$2   & Telesco et al. (1991)      \\
450$\mu$m                      & 16$\pm$2   & Alton et al. (1999)        \\
450$\mu$m                      & 16$\pm$3   & Smith et al. (1991)        \\
450$\mu$m                      & 14$\pm$1   & Hughes et al. (1994)       \\
850$\mu$m                      & 16$\pm$2   & Alton et al. (1999)        \\
1300$\mu$m                     & 18$\pm$4   & Kr{\"u}gel et al. (1990)   \\
2700$\mu$m                     & 17$\pm$2   & Neininger et al. (1998)    \\
3400$\mu$m                     & 18$\pm$4   & Seaquist et al. (1998)     \\
                               &            &                            \\
{\it Spectral lines}$^{c}$     &            &                            \\
Br\,$\gamma$ 2.16$\mu$m        & 10$\pm$2   & Lester et al. (1990)       \\
Br\,$\gamma$ 2.16$\mu$m        & 10$\pm$2   & Larkin et al. (1994)       \\
$[{\rm Ne}$\,{\sc ii}] 12.8$\mu$m& 15$\pm$1 & A \& L (1995)$^{d}$        \\
CO 7--6 0.37\,mm               & 15$\pm$3   & This paper                 \\
$[{\rm C}$\,{\sc i}] 0.6\,mm   & 21$\pm$6   & White et al. (1994)$^{e}$  \\
CO 4--3 0.65\,mm               & 14$\pm$3   & This paper                 \\
CO 3--2 0.9\,mm                & 20$\pm$5   & Tilanus et al. (1991)      \\
$^{13}$CO 3--2 0.9\,mm         & 13$\pm$3   & This paper                 \\
CO 2--1 1.3\,mm                & 24$\pm$3   & Loiseau et al. (1990)      \\
$^{13}$CO 2--1 1.4\,mm$^{f}$   & 15$\pm$5   & Loiseau et al. (1988)      \\
CO 1--0 2.6\,mm                & 26$\pm$2   & Shen \& Lo (1995)          \\
CO 1--0 2.6\,mm                & 25$\pm$3   & Carlstrom (1988)           \\
CO 1--0 2.6\,mm                & 27$\pm$3   & Lo et al. (1987)           \\
CO 1--0 2.6\,mm                & 16$\pm$5   & Nakai et al. (1987)        \\
$^{13}$CO 1--0 2.7\,mm         & 27$\pm$2   & Neininger et al. (1998)    \\
CS 2--1 3.1\,mm                & 26$\pm$3   & Baan et al. (1990)         \\
HCO$^{+}$ 1--0 3.4\,mm$^{g}$   & 20$\pm$2   & Carlstrom (1988)           \\     
HCN 1--0 3.4\,mm               & 24$\pm$3   & Carlstrom (1988)           \\
HI\ 21\,cm                     & 29$\pm$3   & Weliachew et al. (1984)    \\
\hline
\end{tabular}
\end{flushleft}
a) Errors are mainly by-eye-estimates. \\
b) From Fig.\,10 of McLeod et al. (1993) \\
c) Lobe separations refer to maps displaying integrated line intensities. \\
d) Achtermann \& Lacey (1995) \\
e) Since the CO $J$ = 4--3 map of White et al. (1994) lacks the characteristic
   double lobed structure, their [C\,{\sc i}] 492\,GHz map should also be 
   confirmed by independent measurements. \\
f) The given lobe separation refers to Fig.\,2 of Loiseau et al. (1988), 
   displaying integrated line intensities. Peak line temperatures (their 
   Fig.\,1) show a separation of $\sim$25$''$. \\
g) The interferometric HCO$^{+}$ map of Carlstrom (1988; angular resolution: 
   10$''$) shows two pronounced peaks. It is however difficult to identify
   the lobes in the integrated intensity map of Seaquist et al. (1998; 
   angular resolution 3\ffas50 $\times$ 3\ffas25). This may be caused by 
   missing flux in the latter map.
\end{table}
 
Are the two main hotspots observed in the $J$ = 1--0 and 2--1 transitions
identical with those observed in higher excited CO lines? The position angles
(east of north) of the lines connecting the hotspots are slightly smaller in
our submillimeter data (Fig.\,\ref{m82-co7-6-tmb-contour-grey}) than in 
interferometric maps (cf. Lo et al. 1987; Shen \& Lo 1995; Kikumoto et al. 
1987; Neininger et al. 1998). This is likely an error in our data caused 
by measurements at varying hour angles with pointing offsets along the 
azimuth and elevation axes.

More significant is a {\it difference in angular separation}: While the two
main peaks of line emission observed by us are separated by 15$''$$\pm$2$''$,
the interferometric maps (Lo et al. 1987; Shen \& Lo 1995; Neininger et al.
1998) show a separation of 27$''$$\pm$2$''$. This difference is larger than the
positional uncertainties. A larger separation in the low-$J$ transitions is
also supported by filled-aperture measurements of Nakai et al. (1986, 1987) and
Wild et al. (1992) for CO $J$ = 1--0, by Loiseau et al. (1990) for CO $J$ =
2--1, by Loiseau et al. (1988) for $^{13}$CO $J$ = 2--1, and by us for the
$^{12}$CO $J$ = 1--0 and 2--1 lines. The CO $J$ = 3--2 distribution (Tilanus
et al. 1991) shows an intermediate lobe separation.

Which peaks are detected in the CO submillimeter lines? The two main lobes of
CO emission are located almost symmetrically with respect to the kinematical
center, both in lines of low (e.g. Neininger et al. 1998) and high (see
Fig.\,\ref{m82-co-ch-map-grey}) excitation. We can therefore exclude that 
we see the north-eastern lobe and the `compact central core' (see Shen \& Lo 
1995; Neininger et al. 1998) that are separated by $\sim$15$''$. Instead,
{\it the sudden drop of angular separation from 27$''$ to 15$''$ must reveal 
inhomogeneities in the molecular ring that are characterized by changes in 
density and temperature}. In addition to the {\it `high CO excitation
component'} there exists a {\it `low CO excitation component'}, mainly emitting 
in the CO $J$ = 1--0 and 2--1 lines. The transition in lobe separation 
occurs at the $J$ = 3--2 line: In CO $J$ = 3--2 (Tilanus et al. 1991) the 
separation is still $\sim$20$''$. $^{13}$CO $J$ = 3--2 emission with smaller 
optical depths and less photon trapping requires, however, higher excited 
gas so that the lobe separation becomes smaller.

Since lobe separations are often comparable to the angular resolution of the 
particular observation, integrated intensity maps may be misleading as they
easily exhibit structure dominated by the superposition of components with
identical lines of sight but different velocities. To clarify the situation,
we thus present in Fig.\,\ref{m82-pv} position-velocity maps of the CO $J$ = 
4--3 (our HHT map with highest signal-to-noise ratio and best relative
pointing) and CO 2--1 (Wei{\ss}, in preparation) line emission. Beam widths are 
18$''$ and 13$''$, respectively. Surprisingly, {\it the p-v diagrams
do not reproduce the strikingly different lobe separations seen in the 
integrated intensity maps}. This also holds when comparing CO $J$ = 2--1 
with 7--6 emission. {\it This hints at wider line profiles for the higher 
excited CO transitions at the inner edges of the CO $J$ = 1--0 and 2--1 
lobes}. This is corroborated by Fig.\,\ref{m82-pv} that shows, for the 
CO $J$ = 4--3 line, a slightly smaller lobe separation {\it and} a significantly
larger full-width-to-half-power line width towards the inner edge of the 
south-western lobe.
 
\subsection{Other tracers of the interstellar medium}

Table\ \ref{lobes} displays lobe separations determined by various tracers of
atomic line, molecular line, or dust continuum emission. There are three
preferred angular distances: $\sim$26$''$, 15$''$, and 10$''$. Spectral lines
of low excitation show two maxima at an angular distance of $\sim$26$''$.
Highly excited lines and the far-infrared and mm-wave continuum show a lobe
separation of $\sim$15$''$. The continuum in the near and mid infrared and the
Br\,$\gamma$ line show an angular spacing of $\sim$10$''$.

Since the dynamical center of the galaxy is located not too far from the
mid point of the line connecting the lobes, these data show evidence for 
an even more complex structure than indicated by the CO data alone.
The hot dust and Br\,$\gamma$ emission from the inner parts of the ring 
may trace the most recent star formation activity as it propagates 
outwards into the molecular lobes located at larger galactocentric distances 
(Alton et al. 1999).

Since the far infrared and submillimeter continuum are associated with the
central portion of the ring, this is the location where column densities of 
the cool dense gas must be largest. 

\section{Excitation analysis}

The $J$ = 7--6 transition is the CO line with highest rotational quantum
number so far observed in M\,82. Measurements of this line widen the range of
excitation accessible by CO data considerably. To analyse these excitation
conditions, we have performed radiative transfer calculations using a Large
Velocity Gradient (LVG) model describing a cloud of spherical geometry (see
Appendix A).

\subsection{Physical parameters: Our data}

The three submillimeter CO transitions measured by us arise from the 
{\it high excitation component}, show a similar spatial distribution, 
and can therefore be used simultaneously in a radiative transfer
analysis. To obtain true line intensity ratios, the CO 4--3 and 7--6 spectra
were smoothed to the angular resolution of the $^{13}$CO 3--2 data. Table\
\ref{tab:intensities} displays these ratios for a beamwidth of 22$''$, assuming
gaussian beamshapes. As indicated by the spatial distributions (see
Fig.\,\ref{m82-co7-6-tmb-contour-grey}), line ratios are similar toward the 
lobes and the central beam. CO 7--6/4--3 (integrated) line intensity ratios 
are at the order of 0.3, while $^{12}$CO 4--3/$^{13}$CO 3--2 ratios are 7 -- 
10 (see Table\ \ref{tab:intensities}).

In Appendix A (Fig.\,\ref{m82-lvg-twelve}) we plot our LVG line
intensity ratios as a function of density, kinetic temperature,
$^{12}$CO/$^{13}$CO ratio, and CO `abundance' (i.e., $X$(CO)/grad\,$V$ with
$X$(CO) = [CO]/[H$_2$]). For the line parameters given in Table\ 
\ref{tab:intensities}, reasonable solutions can be found for small CO 
abundances ($X$(CO)/grad\,$V$ $\sim$ 10$^{-6}$\,pc/\kms) and high 
$^{12}$CO/$^{13}$CO ratios ($\sim$ 75). High CO abundances ($X$(CO)/grad\,$V$ 
$\sim$ 10$^{-3}$\,pc/\kms) and small $^{12}$CO/$^{13}$CO ratios are however 
not providing a realistic solution, because H$_2$ densities would approach 
10$^{2}$\,\percc\ and would thus become prohibitively small. Calculations 
assuming a plane-parallel instead of a spherical cloud geometry would yield 
even smaller densities.

A comparison of linewidths and velocity drifts with the size of the region
suggests grad\,$V$ $\sim$ 1\,\kms/pc. With this value and $X$(CO) $\sim$
10$^{-5}$ -- 10$^{-4}$ (e.g. Blake et al. 1987; Farquhar et al. 1994) we
can exclude a $^{12}$CO/$^{13}$CO abundance ratio as small as 25 and are
guided to the four diagrams of Fig.\,\ref{m82-lvg-twelve} with log\,$X$(CO)
$\sim$ log\,($X$(CO)/[grad\,$V$/\kms\,pc$^{-1}$]) = $-5$ or $-4$ and
$^{12}$CO/$^{13}$CO = 50 or 75. Our parameters are summarized in 
Table\ \ref{tab:lvg}. Kinetic temperatures are $T_{\rm kin}$ $\sim$ 60 -- 
130\,K, densities \numd\ $\sim$ 10$^{3.3-3.9}$\,\percc, and 
area filling factors $f_{\rm a,22''}$ $\sim$ 0.04 -- 0.07 and $f_{\rm a,15''}$
$\sim$ 0.07 -- 0.11. While the density is well constrained, the 
kinetic temperature is less well determined. Solutions with $T_{\rm kin}$ 
$\ga$ 150\,K are however unlikely because they would require H$_{2}$ densities 
$\la$10$^{3}$\,\percc. Even excluding such extreme solutions, the density of 
the CO emitting gas is small when compared with that of the Orion Hot Core 
(e.g. Schulz et al. 1995; van Dishoeck \& Blake 1998), but {\it column 
densities} are large: With $N$(CO) = 3.08\,10$^{18}$ [$n_{\rm CO}$/\percc] 
[($\Delta V$/grad\,$V$)/pc]\,\cmsq, we obtain the 22$''$ beam averaged 
and cloud averaged column densities displayed in Table\ \ref{tab:lvg}.
For a line-of-sight source size of 350\,pc and $N$(H$_2$)$_{22''}$ = 
10$^{23}$\,\cmsq\ (see Table\ \ref{tab:lvg}), the beam averaged mean 
molecular density is $<$\numd$>_{22''}$ $\sim$ 200\,\percc\ and the 
volume filling factor becomes $f_{\rm v,22''}$ = $<$\numd$>$/\numd\ 
$\sim$ 0.05. Since the ratio $f_{\rm v,22''}$/$f_{\rm a,22''}$ denotes 
the line of sight dimension of the clouds in units of the beam size, 
we obtain with $r_{\rm cloud}$ = 0.5\,tg\,22$''$ $D_{\rm pc}$ 
[$f_{\rm v,22''}$/$f_{\rm a,22''}$] $\sim$ 150\,pc a characteristic 
cloud radius (that will be discussed and revised in Sects.\,6.2 and 6.3). 
The total molecular mass is $M_{\rm mol,22''}$ $\sim$ 1--7\,10$^{8}$\,\solmass.

\begin{figure}
\hspace{0.05cm}
\psfig{figure=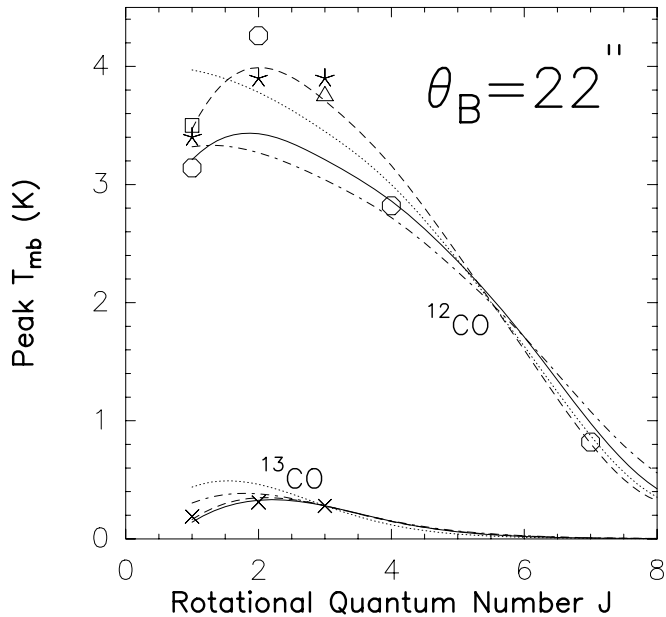,height=9.0cm,angle=-90}
\vspace{0.0cm} 
\caption[]{Peak line temperatures as a function of rotational quantum number
$J$ ($J$ $\rightarrow$ $J$ -- 1) for a beam size of 22$''$ toward the dynamical
center of M\,82. $^{12}$CO: Open square: Wild et al. (1992); stars:
Mauersberger et al. (1999); open triangle: R. Wielebinski (priv. comm.);
circles: this paper. $^{13}$CO (crosses): this paper. Results from radiative 
transfer calculations are given for $^{12}$CO and $^{13}$CO. Solid lines:
$X$(CO)/grad\,$V$ = 10$^{-5}$ and 1.33\,10$^{-7}$\,pc/\kms, \numd\ =
8000\,\percc, $T_{\rm kin}$ = 70\,K; dashed lines: $X$(CO)/grad\,$V$ =
10$^{-5}$ and 2\,10$^{-7}$\,pc/\kms, \numd\ = 5000\,\percc, $T_{\rm kin}$ =
100\,K; dotted lines: $X$(CO)/grad\,$V$ = 10$^{-4}$ and
1.33\,10$^{-6}$\,pc/\kms, \numd\ = 2000\,\percc, $T_{\rm kin}$ = 80\,K;
dash-dotted lines: $X$(CO)/grad\,$V$ = 10$^{-4}$ and 2\,10$^{-6}$\,pc/\kms,
\numd\ = 2000\,\percc, $T_{\rm kin}$ = 130\,K.
}
\label{m82-tmb-22}
\end{figure}

\begin{figure}
\hspace{0.05cm}
\psfig{figure=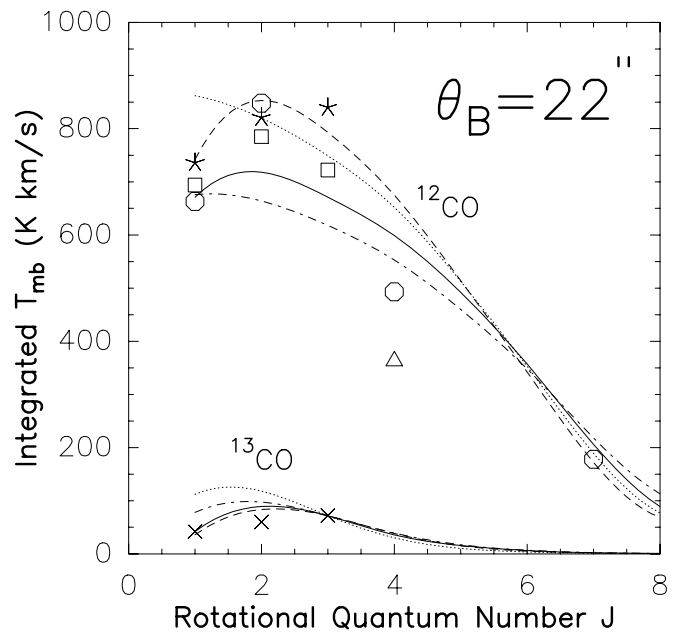,height=9.0cm,angle=-90} 
\vspace{0.0cm}
\caption[]{Integrated line intensities as a function of rotational quantum
number $J$ ($J$ $\rightarrow$ $J$ -- 1) for a beam size of 22$''$ toward the
dynamical center of M\,82. $^{12}$CO: Open squares: Wild et al. (1992); open
triangle: G{\"u}sten et al. (1993); stars: Mauersberger et al. (1999);
circles: this paper. $^{13}$CO (crosses): this paper. While most of our
spatially convolved integrated line intensities were obtained from relatively
extended maps (see Fig.\,\ref{m82-co7-6-tmb-contour-grey}), the convolution of 
our $^{13}$CO $J$ = 2--1 data may be affected by insufficient spatial coverage 
(see Sect.\,2.2).  The CO $J$ = 2--1, 3--2, and 4--3 data from Wild et al. 
(1992) and G{\"u}sten et al. (1993) are based on few observed positions only. 
These cover the ridge of strong CO emission and only allow to estimate 
intensities for the dynamical center of M\,82. To account for the missing 
positions, the resulting integrated intensities were multiplied by 0.85 (the 
factor was deduced from the CO $J$ = 3--2 map of Tilanus et al. 1991). Results 
from radiative transfer calculations are also given (for the parameters, see 
Fig.\,\ref{m82-tmb-22}).
}
\label{m82-int-22} 
\end{figure}

\begin{figure}
\hspace{0.05cm}
\psfig{figure=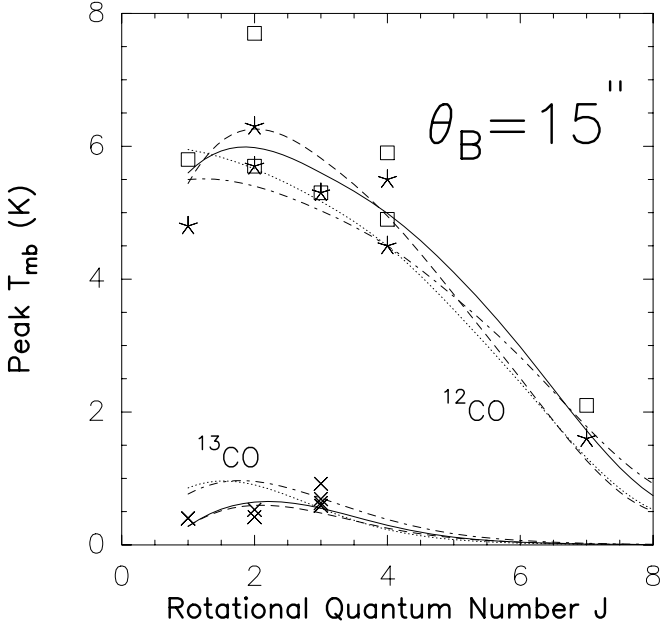,height=9.0cm,angle=-90}
\vspace{0.0cm} 
\caption[]{Peak line temperatures as a function of rotational quantum number
$J$ ($J$ $\rightarrow$ $J$ -- 1) for a beam size of 15$''$ toward the
south-western (squares) and north-eastern (stars) lobe of M\,82. All
temperatures were rescaled for coupling the 15$''$ beam to a source solid angle
(at a given velocity) of $\Omega$ = 100\,arcsec$^{2}$ (e.g. Lord et al. 1996).
$^{12}$CO $J$ = 1--0: Nakai et al. (1987; 16$''$), this paper (22$''$); $J$ =
2--1: Wild et al. (1992; 13$''$), this paper (13$''$); $J$ = 3--2: Tilanus et
al. (1991; 14$''$); $J$ = 4--3: G{\"u}sten et al. (1993; 15$''$), this paper
(18$''$); $J$ = 7--6: this paper (13$''$). $^{13}$CO (crosses): $J$ = 1--0 and
2--1: this paper (22$''$ and 13$''$); $J$ = 3--2: Tilanus et al. (1991;
14$''$), this paper (22$''$). Results from radiative transfer calculations are
also given (for the parameters, see Fig.\,\ref{m82-tmb-22}).
}
\label{m82-tmb-15}
\end{figure}

\begin{table}
\caption[]{\label{tab:lvg} LVG model parameters (compare with 
                           Table\ \ref{tab:pdr})}
\begin{flushleft}
\begin{tabular}{llc}
\hline
                              &                             &        \\
                              &                             &Footnote\\
{\it High excitation comp.}   &                             &        \\
$^{12}$CO/$^{13}$CO           & 50--75                      &   a    \\
$T_{\rm kin}$                 & 60--130\,K                  &        \\
\numd                         & 10$^{3.3 ... 3.9}$\,\percc  &        \\
$f_{\rm a,22''}$              & 0.04 ... 0.07               &   b    \\
$f_{\rm a,15''}$              & 0.07 ... 0.11               &   b    \\
$f_{\rm v,22''}$              & $\sim$0.05                  &   c    \\
$\tau$(CO $J$ = 1--0)         & 0.5 -- 4.5                  &        \\
$\tau$(CO $J$ = 2--1)         & 3.5 -- 15                   &        \\
$\tau$(CO $J$ = 4--3)         & 10 -- 40                    &        \\
$\tau$(CO $J$ = 7--6)         & 6 -- 25                     &        \\
$N$(CO)$_{22''}$              & $\sim$5\,10$^{18}$\,\cmsq   &   d    \\
$N$(H$_2$)$_{22''}$           & $\sim$10$^{23}$\,\cmsq      &   d    \\
$N$(CO)$_{\rm cloud}$         & $\sim$10$^{20}$\,\cmsq      &   e    \\
$N$(H$_2$)$_{\rm cloud}$      & $\sim$10$^{24...25}$\,\cmsq &   e    \\
                              & $\sim$10$^{22}$\,\cmsq/\kms &        \\
$<$\numd$>_{22''}$            & $\sim$200\,\percc           &        \\ 
$r_{\rm cloud}$               & $\sim$150\,pc               &   f    \\
$M_{\rm mol,22''}$            & $\sim$1--7\,10$^{8}$\,\solmass&      \\
                              &                             &        \\
{\it Low excitation comp.}    &                             &        \\
\numd                         & $\sim$10$^{3}$\,\percc      &        \\
\hline
\end{tabular}
\end{flushleft}
a) Abundance ratio \\   
b) $f_{\rm a,22''}$: Area filling factor for the central 22$''$  \\
c) $f_{\rm v,22''}$: Volume filling factor for the central 22$''$  \\
d) $N$(CO)$_{22''}$, $N$(H$_2$)$_{22''}$: 22$''$ beam averaged total column 
   densities \\
e) $N$(CO)$_{\rm cloud}$, $N$(H$_2$)$_{\rm cloud}$: 22$''$ cloud averaged 
   total column densities \\
f) $r_{\rm cloud}$: Characteristic molecular cloud size \\
\end{table}
 
\subsection{Physical parameters: All CO data}

So far, we have only analysed the `warm CO component' of M\,82, exclusively
seen in $^{13}$CO $J$ = 3--2 and higher excited rotational transitions. To
combine these results with data from lower $J$ rotational transitions and to
further elucidate the physical state of the gas, Figs.\,\ref{m82-tmb-22} --
\ref{m82-tmb-15} show (integrated) line intensities as a function of quantum
number $J$. Calibration errors are at the order of $\pm$10\% in the $J$ = 1--0
and 2--1 lines and $\pm$20\% in the higher excited lines. 

Figs.\,\ref{m82-tmb-22} and \ref{m82-int-22} show (integrated) line
temperatures for a 22$''$ beam toward the dynamical center of the galaxy.
Fig.\,\ref{m82-tmb-15} displays line temperatures for a 15$''$ beam toward the
lobes and demonstrates that CO excitation is similar toward the south-western
and north-eastern hotspot. Results from radiative transfer calculations (see
Sect.\,5.1 and Appendix A) are also given. Input parameters correspond to the
four boxes in Fig.\,\ref{m82-lvg-twelve} that provide a promising fit to the
data (log\,($X$(CO)/[grad\,$V$/\kms\,pc$^{-1}$] = $-5$ or $-4$ and
$^{12}$CO/$^{13}$CO = 50 or 75).

Apparently there is a problem with the CO $J$ = 4--3 line: The integrated
intensity (Fig.\,\ref{m82-int-22}) does not allow a reasonable fit, while the
peak intensity (Fig.\,\ref{m82-tmb-22}) is `appropriate'. This is caused by
the narrow lineshape of our 4--3 spectrum (Fig.\,\ref{m82-three-comp-all}).
Compared with other lines, the integrated line intensity is too small, but the
peak line temperature is almost `normal'. The CO $J$ = 4--3 profile, shown
by G{\"u}sten et al. (1993) for the central position, is also weak, both with
respect to its peak and integrated line intensity. The emission from the
lobes, however, fits into the general trend (Fig.\,\ref{m82-tmb-15}). An
interpretation in terms of a diminished lobe separation for CO $J$ $\ge$ 4--3
(see Sect.\,4.1) is not conclusive: The 22$''$ beam centered on the dynamical
core of M\,82 is mainly confined to the inner parts of the lobes and should
not be greatly affected by emission from further out. A CO $J$ = 4--3
deficiency is neither seen in Fig.\,\ref{m82-tmb-15} nor in a corresponding
plot showing integrated intensities for a 15$''$ beam. 

To summarize: Data from the $J$ = 4--3 line are contradictory so that 
convincing evidence for a true anomaly is missing. In spite of differences 
in lobe separation (Sect.\,4.1), the CO data can be reproduced, within 
observational errors, with densities, temperatures, filling factors, 
[CO]/[H$_2$] abundances, and $^{12}$CO/$^{13}$CO isotope ratios
deduced from the three submillimeter transitions mapped by us with the HHT
(Fig.\,\ref{m82-co7-6-tmb-contour-grey}). 

There are few constraints for the {\it low excitation component} that is mainly 
seen in the CO $J$ = 1--0 and 2--1 lines. The column density must be smaller
than for the high excitation component because the latter is more closely 
related to the far infrared and submillimeter continuum from the dust 
(Sect.\,4.2). In view of the remarkable number of `super'-star clusters 
(O'Connell et al. 1995) and supernovae (Kronberg et al. 1985) near the outer 
portions of the ring, cloud temperatures for the low excitation component
should also be $T_{\rm kin}$ $\gg$ 10\,K. For $T_{\rm kin}$ $\sim$ 50\,K, 
LVG densities are at the order of \numd\ $\sim$ 10$^{3}$\,\percc\ or less.

\section{Discussion}

\subsection{A comparison with other LVG simulations}

The most detailed models of CO emission from M\,82 were so far provided by
G{\"u}sten et al. (1993). In order to fit the $^{12}$CO/$^{13}$CO $J$ = 1--0
and 2--1 line intensity ratios then available, they rejected a one component 
LVG scenario and introduced two gas components, one of low (\numd\ $\sim$ 
10$^{3}$\,\percc) and one of high ($\sim$10$^{5}$\,\percc) density. The low 
density component is similar to that proposed by us for the gas mainly 
emitting in the CO $J$ = 1--0 and 2--1 lines (see Sect.\,4.1). Our parameters 
for the high CO excitation component agree to a large extent with those of 
the one component scenario of G{\"u}sten et al. (1993).

For the highly excited gas we thus also find optically thick $^{12}$CO
low-$J$ emission and a high [$^{12}$CO]/[$^{13}$CO] abundance ratio ($\ga$50;
see Table\ \ref{tab:lvg}), that is further supported by an independent chain 
of arguments involving CN and $^{13}$CN data (see Henkel et al. 1998). While 
the LVG model result is shown to be inconclusive in Sect.\,6.3.2, the CN data 
suggest that the [$^{12}$CO]/[$^{13}$CO] ratio is larger than that observed 
in the galactic center region. Likely explanations are radial infall of 
$^{13}$CO deficient gas from the outer parts of the galaxy or a $^{12}$C 
excess in the ejecta from massive stars (e.g. Henkel \& Mauersberger 1993). 
In agreement with G{\"u}sten et al. (1993) we also find that excitation in the 
south-western and north-eastern lobe is similar. Since our new data allow 
us to constrain kinetic temperatures to the high end of those predicted 
by the one component scenario of G{\"u}sten et al. ($T_{\rm kin}$ $\ga$ 
30--70\,K), our densities are at the low end (\numd\ $\la$ 10$^{4}$\,\percc, 
since the product $T_{\rm kin}$ \numd$^{1/2}$ is approximately conserved 
among models simulating optically thick subthermal CO emission). There is 
a remarkable agreement with respect to CO column density, H$_2$ density, 
and molecular gas mass between various studies (cf. Tilanus et al. 1991; 
Wild et al. 1992; G{\"u}sten et al. 1993).

\subsection{Intrinsic inconsistencies of the model}

Our CO data could be reproduced by assuming the presence of two gas
components. Selecting such a `best' model, we could discriminate between
previously proposed scenarios and could constrain cloud conditions giving
rise to highly excited CO emission to slightly higher temperatures and 
smaller densities than previously suggested. On a deeper level, however, 
there remain problems. While calculated CO column densities (Sect.\,5.1) 
appear to be correct (see e.g. Fig.\,2 of Lo et al. 1987 and Fig.\,3 of 
Smith et al. 1991), an obvious puzzle is the large volume filling factor 
($f_{\rm v,22''}$ $\sim$ 0.05) that is comparable to the area filling 
factor (Sect.\,5.1). This forced us to postulate a characteristic cloud 
radius ($r_{\rm cloud}$ $\sim$ 150\,pc) that encompasses a large part of 
the studied volume. Such a large cloud radius is inconsistent with the 
expectation of similar scale lengths along the three dimensions 
($f_{\rm v,22''}$ $\sim$ $f_{\rm a,22''}^{3/2}$) and with the {\it
spatial fine structure} deduced from CO (e.g. Shen \& Lo 1995; Neininger et
al. 1998), high density tracers like HCN and HCO$^{+}$ (e.g. Brouillet \&
Schilke 1993; Paglione et al. 1997; Seaquist et al. 1998), and infrared fine
structure lines (e.g. Lugten et al. 1986; Lord et al. 1996).

Another problem is the {\it density of the gas} observed. A density of \numd\
$\la$ 10$^{4}$\,\percc\ is small when compared to the prototypical `hot core'
associated with the Orion nebula, violating a sometimes noted similarity
between these tiny, highly obscured galactic star forming regions and the more
extended starbursts studied in external galaxies (e.g. Lo et al. 1987; Wolfire
et al. 1990). More seriously, the molecular density determined by us for the
starburst in M\,82 is smaller than most theoretical studies and observational
data permit: Assuming `reasonable' density stucture (\numd\ $\propto$
$r^{-1...-2}$) and accounting for the intense UV field, Brouillet \& Schilke
(1993) find that molecular clouds with densities less than a few times
10$^{4}$\,\percc\ should not exist in the central region of M\,82. For the
transition region between the atomic and molecular gas they propose a density
of 10$^{4-5}$\,\percc\ which disagrees with the range of densities deduced from
our LVG analysis in Sect.\,5.1 (for cloud stability against tidal stress, see
Appendix B).

Studies of high density tracers (e.g. CS, HCN, and HCO$^{+}$) commonly reveal
densities \numd\ $\ga$ 10$^{4}$\,\percc\ (e.g. Mauersberger \& Henkel 1989;
Brouillet \& Schilke 1993; Paglione et al. 1997; Seaquist et al. 1998), even
for the low excitation component (compare Mauersberger \& Henkel 1989 with 
Baan et al. 1990). In view of $T_{\rm dust}$ = 48\,K (Hughes et al. 1994; 
Colbert et al. 1999), radiative excitation by infrared photons should not 
significantly alter the density estimates (see Carrol \& Goldsmith 1981 and
Appendix C). Millimeter wave recombination lines indicate the presence of an 
ionized component with low filling factor and high electron density 
($n_{\rm e}$ $>$ 10$^{4.5}$\,\percc; Seaquist et al. 1996). This gas may be 
associated with a population of (ultra)compact HII regions or with shock 
ionized dense molecular material. Far infrared fine-structure lines, if 
tracing the interface between the molecular and the ionized gas, indicate 
densities of 10$^{3.3-4.0}$\,\percc\ (e.g. Lugten et al. 1986; Wolfire et 
al. 1990; Lord et al. 1996; Stutzki et al. 1997; Colbert et al. 1999) that 
may agree too well with those derived by us for CO (Sect.\,5.1).

An interesting aspect is also provided by {\it kinetic temperature} estimates:
With $T_{\rm kin}$ $\sim$ 20--60\,K (Seaquist et al. 1998; this temperature is
consistent with the estimated cosmic ray flux; see V{\"o}lk et al. 1989,
Suchkov et al. 1993, and Fig.\,2 of Farquhar et al. 1994) the temperature
of the dense molecular gas appears to agree fairly well with that of the dust
($T_{\rm dust}$ = 48\,K; Hughes et al. 1994; Colbert 1999). This is further
supported by an apparent lack of CH$_3$OH and SiO emission, two tracers of high
temperature gas that are easily seen in other nearby starburst galaxies
(Henkel et al. 1991; Mauersberger \& Henkel 1993). The surrounding neutral and
ionized layers have larger temperatures, at the order of 50--100 to 200\,K
(e.g. Lugten et al. 1986; Wolfire et al. 1990; Lord et al. 1996; Colbert et
al. 1999). An analysis of [C\,{\sc i}] 492 and 809\,GHz emission from the
south-western lobe of M\,82 (Stutzki et al. 1997) shows a particularly
striking similarity in density, temperature, and area (but not volume) filling
factor with our LVG parameters derived for CO (Sect.\,5.1; the [C\,{\sc i}]/CO
abundance ratio then becomes $\sim$0.4). It seems as if CO were an integral
part of the dense atomic gas layers of M\,82.

\subsection{A PDR model for the CO emission from M\,82}

\subsubsection{General aspects}

To resolve inconsistencies related to CO fine scale structure, H$_2$ density,
and kinetic temperature (Sect.\,6.2), we note that a significant fraction of
the molecular gas in the Milky Way lies in photon dominated regions (PDRs; see
Hollenbach \& Tielens 1997). A PDR model is also successfully applied to
simultaneously explain CO and [C{\sc ii}] line intensities in the central 
region of the late-type spiral IC\,342 that is not a starburst galaxy but 
that is believed to be a face-on `mirror image' of the galactic center region
(Schulz et al. in preparation). 

In the central region of M\,82 with its high UV flux, the bulk of the 
interstellar line radiation likely arises from PDRs. As we have seen 
(Sect.\,6.2), the coolest gas component is observed toward the dense 
cloud cores, a situation that is consistent with the inside-out temperature 
gradient expected in the case of PDRs. While CO excitation by infrared 
radiation from the dust can be neglected (see Appendix C), the impact of 
UV photons may strongly affect rotational level populations and measured 
CO line intensities in regions with high UV flux. Estimating the far-UV 
flux (6.0 -- 13.6\,eV) in the starburst region, Wolfire et al. (1990), 
Stacey et al. (1991), Lord et al. (1996), and Colbert et al. (1999) find 
$\chi$ $\sim$ 10$^{2.8-3.9}$ ($\chi$: incident far-UV flux in units of 
the local galactic flux, 1.6\,10$^{-3}$\,erg\,s$^{-1}$\,\cmsq). Such a 
high value leads to outer cloud layers that are predominantly heated by 
collisions with electrons photoejected from dust grains or through 
collisional deexcitation of vibrationally excited, UV-pumped H$_2$.

Exploring effects of finite cloud size for clumps with plane-parallel
geometry, K{\"o}ster et al. (1994) presented comprehensive computations of CO
rotational line intensities from PDRs. In contrast to the constant temperature
LVG treatment presented in Sect.\,5, PDR models account for strong kinetic
temperature gradients. The low-$J$ lines typically arise from deeper inside
the clouds than the mid-$J$ lines observed by us (Sect.\,3.1). Reproducing
observed line ratios of optically thick $^{12}$CO emission, PDR model
densities are larger than those derived from a one temperature, one density
LVG model. PDR line intensity ratios only surpass unity if the higher-$J$
line has a critical density (for the values, see Sect.\,1) that is smaller
than the actual gas density. Then, both lines are approximately thermalized but
the higher $J$-line is emitted from warmer layers further out. For densities
at the order of 5\,10$^{3}$\,\percc\ (the density deduced from our one
component LVG model in Sect.\,5.1), $^{12}$CO PDR line temperatures rapidly
decrease with rotational quantum number $J$. For $\chi$ $\sim$ 10$^{3}$ and
$N$(H$_2$) $\sim$ 10$^{22}$\,\cmsq/\kms\ (Sect.\,5.1), densities are then at
the order of 10$^{4}$ and 10$^{5}$\,\percc\ for the low and high excitation
components, respectively.

For the gas with {\it low CO excitation}, a density of \numd\ $\sim$
10$^{4}$\,\percc\ is sufficiently high to explain the detection of CS $J$ =
2--1 emission (e.g. Mauersberger \& Henkel 1989; Baan et al. 1990). A density
of \numd\ $\sim$ 10$^{5}$\,\percc\ in the {\it high excitation} region would 
fulfill all theoretical (Brouillet \& Schilke 1993) and observational density 
requirements outlined in Sect.\,6.2. Furthermore, the volume filling factor 
$f_{\rm v,22''}$ = $<$\numd$>_{22''}$/\numd\ would drop by one to two orders 
of magnitude below the value estimated in Sect.\,5.1. If the area filling 
factor is not drastically altered, this yields reasonable molecular cloud 
radii $\la$10\,pc (see Sect.\,5.1 and Lugten et al. 1986; Wolfire et al. 
1990; Brouillet \& Schilke 1993; Shen \& Lo 1995; Seaquist et al. 1996; 
Stutzki et al. 1997). The apparently intermediate temperature of the CO 
emitting gas between those of the dust and the dense atomic medium is 
naturally explained by mid-$J$ CO emission predominantly arising in the 
heated surface layers of dense molecular clouds.

To summarize: PDR simulations of $^{12}$CO emission remove inconsistencies
related to spatial fine scale structure, density, and kinetic temperature. 
So far published PDR results fail however when $^{13}$CO is also considered. 
Calculated $^{12}$CO/$^{13}$CO line intensity ratios are much smaller than 
the ratios observed (see Table\ \ref{tab:intensities} and K{\"o}ster et al. 
1994) and a more detailed numerical analysis is therefore needed.

\subsubsection{PDR model calculations}

For a numerical approach we note that the beam averaged column density
($N$(H$_2$)$_{22''}$ $\sim$ 10$^{23}$\,\cmsq) and the beam averaged density
($<$\numd$>_{22''}$ $\sim$ 200\,\percc) are observationally determined
(Sects.\,5.1 and 6.2) and do not depend on the choice of the excitation model
if a significant fraction of the dust is associated with molecular clouds. 
Complementing these boundary conditions, we obtain (cf. Sect.\,5.1)
\begin{equation}
\frac{r_{\rm cloud}}{\rm [pc]}\ \sim\ 175\
                \frac{200\,{\rm cm^{-3}}/n\,({\rm H_2})_{\rm PDR}}
                     {T_{\rm CO,observed}/T_{\rm CO,PDR}}
\end{equation}
and
\begin{equation}
\frac{N({\rm H_2})_{\rm PDR}}{\rm [cm^{-2}\,km\,s^{-1}]}\ \sim\ 
                 \frac{10^{23}/300}{T_{\rm CO,observed}/T_{\rm CO,PDR}}
\end{equation}
with 200\,\percc/\numd$_{\rm PDR}$ denoting the volume filling factor, $T_{\rm
CO,observed}/T_{\rm CO,PDR}$ being the area filling factor, \numd$_{\rm PDR}$
giving the average density, and $N$(H$_2$)$_{\rm PDR}$ representing the
average column density of an individual clump per \kms\ (300\,\kms\ is the
total CO linewidth of the nuclear region of M\,82; $N$(H$_2$)$_{\rm PDR}$ =
(4/3)\,$\cdot$\,$r_{\rm cloud}$\,$\cdot$\,\numd$_{\rm PDR}$ and \numd$_{\rm
PDR-cloud-surface}$ = 0.5\,$\cdot$\,\numd$_{\rm PDR}$\ for the assumed 
cloud geometry and density structure, see below).

\begin{figure}
\hspace{-0.20cm}
\psfig{figure=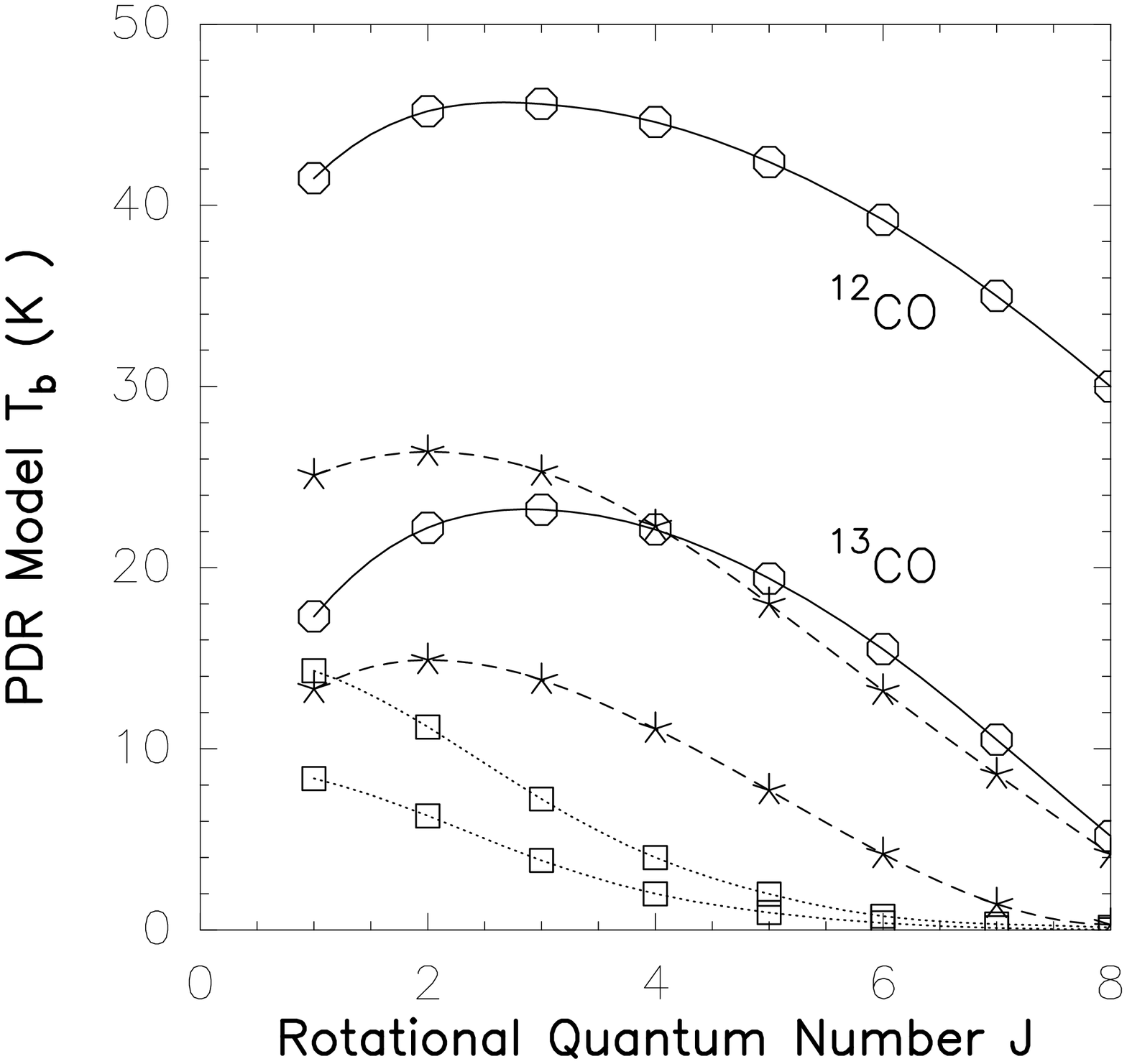,height=9.0cm,angle=-90}
\vspace{0.0cm} \caption[]{PDR model peak line temperatures as a function of
rotational quantum number $J$ ($J$ $\rightarrow$ $J$ -- 1) for clouds with
high column density. Dotted lines with open squares: \numd\ =
2\,10$^{4}$\,\percc, the curve with high $T_{\rm b}$ displays $^{12}$CO, the
lower one $^{13}$CO; dashed lines with asterisks: \numd\ =
2\,10$^{5}$\,\percc; solid lines with open circles: \numd\ =
2\,10$^{6}$\,\percc. For all models, $N$(H$_2$) = 2\,10$^{22}$\,\cmsq,
implying cloud radii of 7.5\,10$^{17}$, 7.5\,10$^{16}$, and
7.5\,10$^{15}$\,cm, respectively. Assumed linewidth of an individual cloud and
its carbon isotope ratio: 1.2\,\kms\ and 67. For the definition of \numd\ =
\numd$_{\rm PDR}$, see Sect.\,6.3.2.
}
\label{m82-pdr1}
\end{figure}

Since embedded clumps of dense gas may more often be spherical than
plane-parallel, spherical clouds with a power law density distribution
($n$($r$) $\sim$ $r^{-1.5}$) were modeled with the dust being heated by the
external UV radiation and intrinsic infrared emission (cf. St{\"o}rzer et al.
1996; 2000). While we can reproduce the observed relative intensities of the
various rotational CO transitions with a density at the order of \numd\ $\sim$
10$^{5}$\,\percc\ (see Fig.\,\ref{m82-pdr1} and Sect.\,6.3.1), $r_{\rm cloud}$
and $N$(H$_2$) are inconsistent with Eqs.\,1 and 2. Furthermore, computed
$^{12}$CO/$^{13}$CO line intensity ratios show little dependence on cloud
structure and remain with characteristic values of 2--4 (see also Gierens et
al. 1992 for $^{12}$C/$^{13}$C = 40; K{\"o}ster et al. 1994 and St{\"o}rzer et
al. 2000 for $^{12}$C/$^{13}$C = 67) much smaller than the observed ratios of
10--15 (see Table\ \ref{tab:intensities}). Varying $\chi$ over the permitted 
range (10$^{3.3\pm 0.4}$) does not change these line intensity ratios 
significantly. Drastically increasing the carbon isotope ratio leads to 
$^{12}$C/$^{13}$C $>$ 100 which is inconsistent with data from the solar 
system, the Milky Way, and the Magellanic Clouds (e.g. Wilson \& Rood 1994; 
Chin et al. 1999).

Other parameters that can be varied are the density and column density of the
far-UV irradiated cloud. The $^{12}$CO lines are formed in warm layers near
the surface whereas $^{13}$CO lines are predominantly emitted from cooler
regions deeper inside. With the column density being fixed, a higher density
leads to a moderate increase in the $^{12}$CO/$^{13}$CO line intensity ratio
(see Fig.\,\ref{m82-pdr1}), because $^{12}$CO is then emitted from CO layers
closer to the heated cloud surface. To reproduce observed $^{12}$CO/$^{13}$CO
line ratios in this way requires, however, densities $\ga$10$^{8}$\,\percc.

\begin{figure}
\hspace{-0.20cm}
\psfig{figure=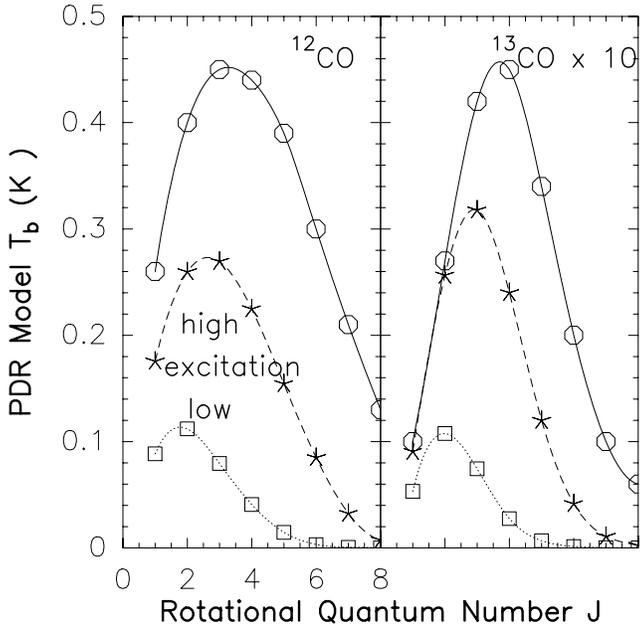,height=9.0cm,angle=-90}
\vspace{0.0cm} 
\caption[]{PDR model peak line temperatures as a function of rotational
quantum number $J$ ($J$ $\rightarrow$ $J$ -- 1) for low column density clouds,
reproducing line intensity ratios for the `high' and `low' CO excitation 
components. Dotted lines with open squares (`low excitation'): \numd\ = 
10$^{3}$\,\percc, $r_{\rm cloud}$ = 4.5\,10$^{17}$\,cm, and $N$(H$_2$) = 
6\,10$^{20}$\,\cmsq; dashed lines with asterisks (`high excitation'): \numd\ 
= 5\,10$^{3}$\,\percc, $r_{\rm cloud}$ = 7.5\,10$^{16}$\,cm, and $N$(H$_2$) 
= 5\,10$^{20}$\,\cmsq; solid lines with open circles: \numd\ = 
2\,10$^{4}$\,\percc, $r_{\rm cloud}$ = 1.5\,10$^{16}$\,cm, and $N$(H$_2$) = 
4\,10$^{20}$\,\cmsq. Assumed linewidth of an individual cloud and its carbon 
isotope ratio: 1.2\,\kms\ and 67, respectively. For the definition of \numd\ 
= \numd$_{\rm PDR}$, see Sect.\,6.3.2. }
\label{m82-pdr2}
\end{figure}

More important are variations in column density: A spherical UV illuminated
cloud is effectively smaller in $^{13}$CO than in $^{12}$CO and this 
difference becomes more pronounced when $N$(H$_2$) decreases (see St{\"o}rzer 
et al. 2000). In the extreme case of totally photodissociated $^{13}$CO, some 
$^{12}$CO may still exist in the clump so that, in principle, 
$^{12}$CO/$^{13}$CO line intensity ratios up to very large values can be 
reproduced. Once the $^{13}$CO abundance gets very small, $^{12}$CO also tends 
to become optically thin. In Fig.\,\ref{m82-pdr2} we show expected line 
temperatures in this optically thin regime, with \numd\ = 5\,10$^{3}$\,\percc\ 
successfully reproducing measured $^{12}$CO line intensity ratios for the {\it 
high CO excitation component}. $^{12}$CO/$^{13}$CO = 19, 10, 8.5, 9.4, 13, 20, 
and 30 for $J$ = 1--0 ... 7--6, respectively. This is compatible with observed 
ratios of 15, 10, and 9 for the three lowest rotational transitions. For the 
{\it low excitation component}, \numd\ $\sim$ 10$^{3}$\percc; 
Fig.\,\ref{m82-pdr2} shows a fit with $^{12}$CO/$^{13}$CO = 17, 10, and 11 for 
$J$ = 1--0 to 3--2.

\begin{table}
\caption[]{\label{tab:pdr} PDR model parameters }
\begin{flushleft}
\begin{tabular}{llc}
\hline
                             &                               &         \\
                             &                               &Footnote \\
{\it High excitation comp.}  &                               &         \\
$^{12}$CO/$^{13}$CO          & undetermined                  &   a     \\
\numd                        & $\sim$10$^{3.7}$\,\percc      &         \\
$f_{\rm a,22''}$             & $\sim$15--200                 &   b     \\
$f_{\rm v,22''}$             & $\sim$ 0.05                   &   c     \\
$N$(CO)$_{22''}$             & $\sim$5\,10$^{18}$\,\cmsq     &   d     \\
$N$(H$_2$)$_{22''}$          & $\sim$10$^{23}$\,\cmsq        &   d     \\
$N$(H$_2$)$_{\rm cloud}$     & $\sim$5\,10$^{20}$\,\cmsq/\kms&   e     \\

$<$\numd$>_{22''}$           & $\sim$200\,\percc             &         \\ 
$r_{\rm cloud}$              & $\sim$0.025\,pc               &   f     \\
$M_{\rm mol,22''}$           & $\sim$1--7\,10$^{8}$\,\solmass&         \\
                             &                               &         \\
{\it Low excitation comp.}   &                               &         \\
\numd                        & $\sim$10$^{3}$\,\percc        &         \\
$f_{\rm a,22''}$             & $\sim$15                      &   b     \\
$r_{\rm cloud}$              & $\sim$0.15\,pc                &   f     \\
\hline
\end{tabular}
\end{flushleft}
a) $^{12}$CO/$^{13}$CO: abundance ratio \\
b) $f_{\rm a,22''}$: Area filling factor for the central 22$''$; the 
   lower value is derived from a comparison of modelled and observed
   line intensities, the higher value results from a comparison of 
   total 22$''$ beam averaged and individual cloud column density.  \\
c) $f_{\rm v,22''}$: Volume filling factor for the central 22$''$ (see 
   Sect.\,5.1) \\
d) $N$(CO)$_{22''}$, $N$(H$_2$)$_{22''}$: 22$''$ beam averaged total
   column density \\
e) $N$(H$_2$)$_{\rm cloud}$: Column density of an individual cloud \\
f) $r_{\rm cloud}$: Characteristic molecular cloud size \\
\end{table}
 
It is remarkable that {\it PDR model densities are similar to those derived
with our LVG simulation}, contradicting theoretical predictions of minimum
densities in excess of 10$^{4}$\,\percc\ (Brouillet \& Schilke 1993). {\it
Column densities} ($N$(H$_2$) $\sim$ 5\,10$^{20}$\,\cmsq/\,\kms) and {\it line
temperatures of individual cloudlets} are, however, drastically different. An
average cloudlet shows optically thin, not optically thick CO line emission.
Line temperatures of individual clumps are much smaller than the observed
average over the inner 300\,pc, so that the {\it PDR area filling factor} 
$f_{\rm a}$ is not $\ll$1 as in the LVG approximation but $\gg$1. Thus not a 
higher density (as anticipated in Sect.\,6.3.1) but a higher $f_{\rm a}$ 
leads to plausible $r_{\rm cloud}$ values in the sub-parsec range (see Eq.\,1). Both $f_{\rm a}$ $\gg$ 1 and $r_{\rm cloud}$ $<$ 1\,pc are consistent with 
early CO studies, favoring CO lines of low optical depth. These conclusions are
supported by observations of fine-structure lines in the infrared (e.g.
Olofsson \& Rydbeck 1984; Lugten et al. 1986; Wolfire et al. 1990; Brouillet
\& Schilke 1993; Schilke et al. 1993; Lord et al. 1996). In view of our PDR
simulations, the high density ($>$10$^{4.5}$\,\percc) ionized gas found by
Seaquist et al. (1996) is likely related to (ultra)compact HII regions and 
not to evaporating dense molecular clouds. 

Obviously, our study does not imply that the core of M\,82 does not contain
regions of high density and column density. The {\it bulk of the CO emission},
however, {\it must arise from a warm low density interclump medium}. The gas
may be barely dense enough to avoid tidal disruption (see Appendix B). For
typical cloud cores in the galactic disk, Eq.\,5 of Larson (1981) predicts
a column density of $N$(H$_2$) $\sim$ 10$^{22}$\,\cmsq. With $N$(H$_2$) $\sim$ 
5\,10$^{20}$\,\cmsq/\kms\ (Table\ \ref{tab:pdr}) this infers a linewidth of 
20\,\kms. The relations defined for approximately virialized clouds by Larson 
(1981) then yield cloud sizes well in excess of 100\,pc, in strong 
contradiction with our PDR model. {\it The bulk of the CO emission therefore 
arises from gas that may not be virialized}. We conclude that {\it CO is 
tracing a different gas component than molecular high density tracers} like 
CS or HCN. The {\it dominance of CO emission from a diffuse medium} with 
cloud fragments of low column density per \kms, $X$ = $N$(H$_2$)/$I_{\rm CO}$ 
conversion ratio, and CO intensity explains why HCN is a better tracer of star 
formation and infrared luminosity (Solomon et al. 1992a). It {\it also 
explains why [C\,{\sc i}]/CO abundance ratios are higher than those observed 
in spatially confined star forming regions of the galactic disk} (see also 
Sect.\,6.2, Schilke et al. 1993, and White et al. 1994).

An argument against our scenario could be that Eqs.\,1 and 2 only hold
approximately: The denominator $T_{\rm CO, observed}/T_{\rm CO, PDR}$ is too
small by an order of magnitude to yield the proper $r_{\rm cloud}$ value
assumed in the PDR approximation, while it is too large by the same amount to
provide the proper $N$(H$_2$)$_{\rm PDR}$ value. In order to check whether our
scenario is plausible, we therefore extend our analysis to other galaxies.

\subsection{Other galaxies}

\subsubsection{The Milky Way and other `quiescent' galaxies}

Studying the physics of the molecular gas in the galactic center region,
Dahmen et al. (1998) compared $^{12}$CO and C$^{18}$O $J$ = 1--0 data obtained
with a linear resolution of $\sim$22\,pc (8$'$). Lineshapes, spatial
distributions, and line ratios indicate the presence of an extended diffuse
gas component that is not apparent in studies focusing on individual cloud
cores (see also Oka et al. 1998a). A significant fraction of the gas is not
virialized. Higher resolution data (e.g. H{\"u}ttemeister et al. 1998) reveal 
that the dense molecular gas (\numd\ $\ga$ 10$^{4}$\,\percc) is relatively 
cool ($T_{\rm kin}$ $\sim$ 25\,K) in comparison to that at lower density 
($\sim$ 100\,K at a few 10$^{3}$\,\percc). Apparently, a warm diffuse 
molecular medium is not only ubiquitous in the starburst galaxy M\,82 but 
also dominates the CO emission in the more quiescent central region of the 
Milky Way (for extended CO maps of a galactic disk star forming region, see 
Wilson et al. 1999). This also holds for the nuclear regions of IC\,342
(Schulz et al. in preparation) and NGC\,7331 (Israel \& Baas 1999), another two
galaxies with $L_{\rm FIR}$ $\sim$ 10$^{10}$\,\solum. Like M\,82, NGC\,7331 
shows a warmer inner and cooler outer dust ring and a molecular ring 
that appears to be related to its cool dust component.

Shocks and cloud-cloud collisions induced by the presence of bars (e.g. 
Achtermann \& Lacey 1995; Morris \& Serabyn 1996; Fux 1997, 1999; 
H{\"u}ttemeister et al. 1998), tidal disruption of clouds near the
center (G{\"u}sten 1989), a high gas pressure (e.g. Helfer \& Blitz 1997) that
can help to keep the gas molecular, and a high stellar density that can
affect molecular cloud dynamics (e.g. Mauersberger et al. 1996b; Oka et al.
1998b) may all contribute to the disintegration of molecular clouds and to the 
synthesis of an extended warm molecular spray consisting of low column density 
cloud fragments.

\subsubsection{Nearby starburst galaxies}

So far, M\,82 is the only starburst galaxy where we could show that a CO-LVG
excitation analysis (inferring $f_{\rm v}$ $\sim$ $f_{\rm a}$; see Sect.\,5.1) 
does not lead to results that are free of inconsistencies. The lack of CO 
$J$ = 7--6 data does not constrain LVG parameters sufficiently to search for 
a similar discrepancy in the other two nearby starburst galaxies NGC\,253 
and NGC\,4945. This implies that previous studies did not provide sufficient
motivation to replace LVG excitation analyses by PDR scenarios. 

Towards NGC\,253, M\,82, and NGC\,4945 $^{12}$CO and $^{12}$CO/$^{13}$CO line 
intensity ratios are similar. This even holds for absolute integrated line 
intensities (within 30\%) for a 22--24$''$ beam size (compare Table\ 
\ref{tab:intensities} with Mauersberger et al. 1996a,b; Harrison et al. 1999).
Not surprisingly, published LVG densities and area filling factors for
NGC\,253 and NGC\,4945 match those for M\,82 (Henkel et al. 1994; Mauersberger 
et al. 1996a,b; Harrison et al. 1999). Following the procedure outlined in 
Sect.\,5.1, average densities along the line-of-sight ($<$\numd$>$ $\sim$ 
100\,\percc) yield volume filling factors at the order of $f_{\rm v,22''}$ = 
$<$\numd$>$/\numd\ $\sim$ 0.02. Since volume filling factors seem to be 
slightly smaller than area filling factors, average cloud sizes become
$\sim$30\,pc (for the Milky Way, see Oka et al. 1998b), not enough to 
request the use of PDR models.

Because of the similarity of CO line strengths and line ratios and since
high [C\,{\sc i}]/CO ratios are observed toward both M\,82 and NGC\,253 
(Schilke et al. 1993; White et al. 1994; Harrison et al. 1995; Israel
et al. 1995; Stutzki et al. 1997) we feel nevertheless that PDR 
simulations relating the 
bulk of the CO emission to a warm diffuse molecular medium are relevant not 
only to M\,82 but to NGC\,253 and NGC\,4945 as well.

\subsubsection{Mergers}

Active disk galaxies exhibit $^{12}$CO/$^{13}$CO ratios $\sim$ 10 (this is
independent of inclination, Hubble Type, and metallicity; Sage \& Isbell
1991). Perturbed gas-rich mergers with infrared luminosities $\ga$
10$^{11}$\,\solum\ tend to show larger values, up to 40--60 (Henkel \&
Mauersberger 1993). This may be caused by excitation effects (see Aalto et al.
1999) or by a deficiency of $^{13}$CO emission since \irlum\ is better 
correlated with $^{12}$CO (Taniguchi \& Ohyama 1998). Inflow of gas from 
the outer disks may provide $^{13}$CO deficient gas and nucleosynthesis 
in shortlived massive stars may further enhance $^{12}$C (Casoli et al. 
1992). While this may help to enhance $^{12}$CO/$^{13}$CO line intensity 
ratios, this is not sufficient to explain measured $^{12}$CO/$^{13}$CO 
values within the context of PDRs. Instead we suggest that the direct 
interaction of two galactic nuclei and their associated disks is even 
more efficient than the presence of a bar to trigger cloud-cloud collisions 
and to create a warm diffuse molecular debris containing a large number of 
small cloud fragments. If such fragments are smaller than in individual 
non-merging galaxies, higher $^{12}$CO/$^{13}$CO line ratios might result. 
While the detailed spatial CO fine-structure may only be revealed by next 
generation mm-wave telescopes (e.g. Downes 1999), this would imply that in 
mergers with $^{12}$CO/$^{13}$CO $\gg$ 10, [C\,{\sc i}]/CO line intensity 
ratios should be as large as or {\it even larger} than in M\,82 and NGC\,253.
For a first detection of [C\,{\sc i}] in a merger, see Gerin \& Philips
(1998).

\subsubsection{Galaxies at high redshifts}

Is it a general property of the integrated spectrum of a starburst that,
beginning with the CO $J$ = 4--3 transition, line intensities (accounting for
beam dilution) gradually decrease with increasing rotational quantum number
$J$? Ground based measurements of nearby galaxies in mid and high-$J$ CO
transitions require exceptional weather conditions. In galaxies of high
redshift, however, many otherwise inaccessible transitions are shifted into
the observable mm- and submm-wavelength bands. While linear resolutions 
remain poor, the bulk of the CO emission from higher excited states should
arise from the nuclear starburst environment. Suitable examples of distant
sources with a number of detected CO transitions are IRAS F\,1024+4724 ($z$ =
2.286), the Cloverleaf quasar ($z$ = 2.558), and BR\,1202--0725 ($z$ = 4.692).
Toward IRAS F\,1024+4724, the $J$ = 3--2 line is stronger than the 4--3 and
6--5 lines; line intensity ratios w.r.t. 3--2 are $\sim$0.75 and 0.6 (Solomon 
et al. 1992b). Toward the Cloverleaf quasar, CO lines are characterized by a 
constant or increasing brightness temperature from 3--2 to 4--3, followed by 
constant or (more likely) decreasing line intensities in the higher $J$ 
transitions (Barvainis et al. 1997). Toward BR\,1202--0725, the $J$ = 7--6 
to 5--4 line intensity ratio is $\sim$0.65 (Omont et al. 1996). Although the 
sample is too small for a reliable statistical analysis, it seems that CO 
line intensity ratios are fairly uniform in starbursts and depend little on 
redshift (age) and temperature ($T_{\rm cmb}$/[K] = 2.73\,(1+$z$)) of the 
microwave background. Unfortunately, $^{12}$CO/$^{13}$CO line intensity 
ratios are not yet known. If they are $\ga$10, densities at the order of 
a few 10$^{3}$\,\percc\ and small column densities 
($\la$10$^{21}$\,\cmsq/\kms) should be a characteristic feature for the 
bulk of the CO emitting cloudlets in all starburst galaxies.

\section{Conclusions}

We have studied millimeter and submillimeter CO line emission up to the $J$ =
7--6 rotational transition toward the central region of the starburst galaxy
M\,82 and obtain the following main results:

\begin{enumerate} \renewcommand{\labelenumi} {(\arabic{enumi})}

\item The spatial structure of the millimeter and submillimeter CO emission is
distinct. While integrated intensity maps suggest that the lobe separation of 
the low-$J$ transitions is $\sim$26$''$, it is $\sim$15$''$ for the mid-$J$ 
transitions. Major-axis position-velocity maps in the CO $J$ = 2--1 and 4--3 
lines show however agreement in the lobe positions. This indicates that,
at the inner edges of the low-$J$ CO lobes, line profiles are wider in the 
higher excited CO transitions. We thus distinguish between a `low' and a `high'
CO excitation component, the latter coinciding with the main source of
millimeter and submillimeter dust emission. 

\item An LVG excitation analysis of CO submillimeter lines leads to internal
inconsistencies. While measured line intensities are reproduced with 
$T_{\rm kin}$ $\sim$ 60 -- 130\,K, \numd\ $\sim$ 10$^{3.3-3.9}$\,\percc, 
cloud averaged column densities $N$(CO)$_{\rm cloud}$ $\sim$ 10$^{20}$ and 
$N$(H$_2$)$_{\rm cloud}$ $\sim$ 10$^{24-25}$\,\cmsq, [$^{12}$CO]/[$^{13}$CO] 
abundance ratios $\ga$50, and a total molecular mass of a few 
10$^{8}$\,\solmass, area filling factors ($f_{\rm a}$ $\sim$ 0.05--0.10) and 
volume filling factors ($f_{\rm v}$ $\sim$ 0.05) are similar. This results
in cloud sizes that do not match their angular scale. On the other hand,
the resulting H$_2$ column density is consistent with that derived from the 
dust continuum at millimeter and submillimeter wavelengths. For the low 
excitation component, densities are $\sim$ 10$^{3}$\percc. 

\item An application of PDR models resolves the inconsistencies of the LVG
calculations. LVG densities, column densities, and total mass are confirmed. 
The bulk of the CO emission arises, however, from a diffuse, low column density
($N$(H$_2$) $\sim$ 5\,10$^{20}$\,\cmsq/\kms) interclump medium with small
$X$ = $N$(H$_2$)/$I_{\rm CO}$ conversion factors, area filling factors $\gg$1,
and sub-parsec cloud sizes. The relations defined by Larson (1981) are not 
fulfilled and the gas may not be virialized. Such a scenario explains why 
CS or HCN are better tracers of global star formation rate and infrared 
luminosity than CO. Our scenario also explains observed high [C\,{\sc i}]/CO 
line intensity ratios, while relative abundances of $^{12}$CO versus $^{13}$CO 
cannot be accurately determined. Higher column density clouds, even accounting 
for variations in far-UV flux and $^{12}$C/$^{13}$C isotope ratios, do not 
reproduce observed $^{12}$CO/$^{13}$CO line intensity ratios $\ga$10. 
Densities are close to the minimum values required for tidal stability in 
the absence of magnetic fields.

\item In regard to $^{12}$CO line intensity ratios, the central region of
M\,82 appears to be representative for the entire family of starburst
galaxies, both at small and at high redshifts. A comparison of the starburst
regions in M\,82 and NGC\,253 demonstrates that this similarity extends
to $^{12}$CO/$^{13}$CO and [C\,{\sc i}]/CO line intensity ratios. The
large $^{12}$CO/$^{13}$CO line intensity ratios ($\gg$10) observed toward
`nearby' mergers prove, however, that differences exist at least w.r.t.
rare CO isotopomers. Galaxy pairs with such high $^{12}$CO/$^{13}$CO line
ratios require the presence of a particularly diffuse highly fragmented low 
column density ISM.

\end{enumerate}

Apparently, dropping the assumption of constant temperature in the CO
excitation model is a necessary step to provide a self-consistent approach to
the physical properties of molecular clouds in the nuclear starburst region of
M\,82. While the use of PDR models is crucial for a better understanding of
the molecular gas phase in a starburst environment, important information is
still missing. Interferometic observations of high density tracers (e.g. CN,
CS, HCN, HNC, N$_2$H$^{+}$), coupled with PDR model calculations including
chemical aspects, are needed to fully understand the spatial morphology,
density distribution, and molecular excitation of this archetypical starburst
complex. An interesting aspect is provided by HCO$^{+}$ $J$ = 1--0 line
emission (Table\ \ref{lobes}). The bulk of this emission might arise from
regions intermediate between those of the low and high CO excitation
component. Since this molecule (as well as N$_2$H$^{+}$) is a sensitive 
tracer of ionization conditions in the dense gas, a detailed knowledge of 
its spatial distribution would be crucial for a better understanding of 
structure and excitation.

So far, models were calculated for the cosmic ray flux of the solar
neighbourhood. A flux enhancement by two to three orders of magnitude (with
all necessary chemical implications) has still to be incorporated into PDR codes
(but see Schilke et al. 1993). Another important quantity is the spatial
distribution of the UV flux. For M\,82, we do not know the variation of the UV
flux as a function of galactocentric radius.

\appendix

\section{The radiative transfer model}

Systematic motions and microturbulence are frequently used as simplifying
assumptions to facilitate treatment of line formation in molecular clouds. For
the large amplitude systematic motions assumed in the Large Velocity Gradient
(LVG) approximation, the source functions are locally defined. 
For the homogeneous velocity field assumed in the microturbulent
approximation, the source functions are generally coupled throughout the cloud
by scattered radiation. The crucial difference between the two approximations
is that between local and non-local excitation and both can be viewed as
limiting cases for the treatment of molecular line formation. In the case of
CO, line intensities determined with a standard LVG model (e.g. Castor 1970;
Scoville \& Solomon 1974) do not differ by more than a factor of three from
models using the microturbulent approach (White 1977; see also Ossenkopf
1997). This is within the uncertainties that can be attributed to cloud
geometry. Furthermore, Wild et al. (1992) found no significant difference
between results from their `clumpy cloud' and standard LVG models. Given the
inhomogeneity of the ISM, the assumption of uniform physical conditions is
crude and only yields average gas properties. In view of the limited quality
of the available data and the lack of information on source morphology and its
fine-scale structure, however, an LVG treatment of the radiative transfer is
an appropriate first step to analyse M\,82.

\begin{figure*}
\hspace{0.9cm}
\psfig{figure=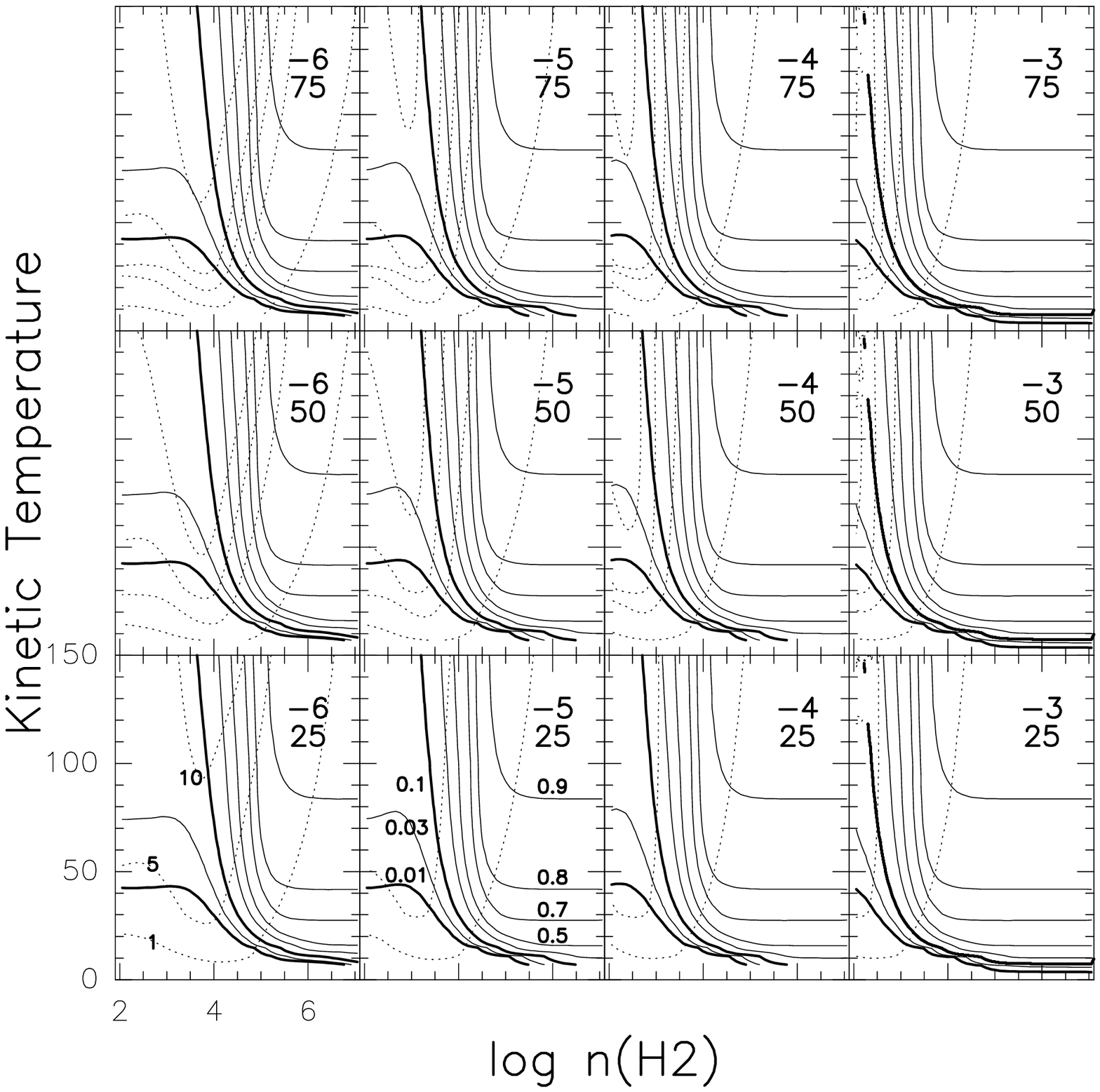,height=14.7cm,angle=-90}
\vspace{0.6cm} 
\caption[A1]{Large Velocity Gradient (LVG) model results for $X$(CO)/grad\,$V$ =
10$^{-6}$, 10$^{-5}$, 10$^{-4}$, and 10$^{-3}$\,pc/\kms\ ($X$(CO) is the
fractional CO abundance parameter) and $^{12}$CO/$^{13}$CO abundance ratios of
25, 50, and 75. The logarithm of $X$(CO)/grad\,$V$ and the $^{12}$CO/$^{13}$CO 
abundance ratio are given for each box in the upper right hand corner. Solid
lines: CO 7--6/4--3 line intensity ratios (contour levels from lower left to
upper right: 0.01, 0.03, 0.1, 0.3, 0.5, 0.7, 0.8, and 0.9). Dotted lines:
$^{12}$CO 4--3/$^{13}$CO 3--2 line intensity ratios (contour levels from lower
right to upper left: 1., 5., 10., 15., and 25.; for $^{12}$CO/$^{13}$CO = 25,
line intensity ratios are $<$15, for $^{12}$CO/$^{13}$CO = 50, line intensity
ratios are $<$25).
}
\label{m82-lvg-twelve}
\end{figure*}

Applying our LVG model to simulate a cloud of spherical geometry, CO 
collisional cross sections were taken from Green \& Chapman (1978). For 
$T_{\rm kin}$ $>$ 100\,K, we assumed collision rates $C_{\rm ij}$ $\propto$ 
$T_{\rm kin}^{1/2}$. Using instead the collision rates recommended by De 
Jong et al. (1975; these include higher temperatures) leads to similar results.
A comparison of $J$=2--1/$J$=1--0 line intensity ratios with those of Castets 
et al. (1990), that were computed with yet another set of collision rates, 
also shows consistency within 20\%.

Fig.\,\ref{m82-lvg-twelve} displays our calculated CO 7--6/4--3 and CO
4--3/$^{13}$CO 3--2 line intensity ratios as a function of density
(10$^{2-7}$\,\percc), kinetic temperature (5--150\,K), $^{12}$CO/$^{13}$CO
abundance ratio (25--75), and CO fractional abundance in terms of
$X$(CO)/grad\,$V$ (10$^{-6 ... -3}$\,pc/\kms; $X$(CO) is the fractional
abundance parameter, grad\,$V$ denotes the velocity gradient in \kms/pc).
The choice of $X$(CO)/grad\,$V$ is motivated by a source size of $\sim$
300\,pc and a velocity range of $\sim$300\,\kms\ (this leads to grad\,$V$ $\sim$
1\,\kms\,pc$^{-1}$), a solar system [C]/[H] abundance ratio of 3.5\,10$^{-4}$
(Grevesse et al. 1994), and the assumption that a significant fraction of the
available carbon ($\ga$10\%) is forming CO molecules. Since molecular clouds 
with significant CO 7--6 emission must be warm, strong CO fractionation in
favor of enhanced $^{13}$C abundances (cf. Watson et al. 1976) can be
excluded. Fig.\,\ref{m82-lvg-twelve} thus presents calculations for 
$^{12}$CO/$^{13}$CO = 25 (the carbon isotope ratio in the galactic center
region; e.g. Wilson \& Rood 1994), $^{12}$CO/$^{13}$CO = 50 (the
$^{12}$C/$^{13}$C ratio in the inner galactic disk; e.g. Henkel et al. 1985),
and $^{12}$CO/$^{13}$CO = 75 (close to the carbon isotope ratio of the local
ISM and the solar system; e.g. Stahl et al. 1989; Stahl \& Wilson 1992).

\section{Cloud stability}

In order for a cloud to be gravitationally bound, it must be sufficiently
dense to withstand the tidal stresses caused by the gravitational potential of
the galaxy. Neglecting rotation, turbulence, and magnetic fields, and
compensating tidal forces by a cloud's own gravity, we can derive a minimum
density that is required for survival in a hostile medium. With $V_{\rm rot}$
$\sim$ 140\,\kms\ at a galactocentric radius of $R$ $\ga$ 75\,pc from the
dynamical center of M\,82 (Neininger et al. 1998) and applying Eq.\,5 of
G{\"u}sten and Downes (1980), we find a limiting cloud density of
$n_{\rm min}$ $\sim$ 10$^{3.2-3.5}$\,[$R$/120\,pc]$^{-2}$\,\percc. $R$ =
120\,pc refers to the molecular lobes of M\,82, located at offsets 
$\pm$7\ffas5 from the dynamical center of the galaxy (see Table \ref{lobes}).
The minimum density $n_{\rm min}$ is near the lower limit of the densities
deduced in Sect.\,5.1 with an LVG model and just matches densities derived in
Sect.\,6.3.2 with a PDR model.

\section{Radiative pumping}

Because the kinetic temperature is constrained by the CO $J$=7--6/$J$=4--3 line
temperature ratios (Table\ \ref{tab:intensities}), collisional excitation to
excited vibrational or electronic levels, $\ga$ 3000\,K above the ground
state, is not effective. Densities and temperatures obtained with
Fig.\,\ref{m82-lvg-twelve} might however be affected by radiative excitation,
that is neglected in our LVG treatment (apart of the 2.7\,K microwave
background). 

To operate efficiently, radiative pumping must be faster than the collisional
rates for rotational excitation in the $v$ = 0 vibrational ground state. With
$<$$\sigma v$$>$ $\sim$ 2\,10$^{-11}$\,cm$^{3}$\,s$^{-1}$ (Green \& Chapman
1978) and a density of 5\,10$^{3}$\,\percc, collision rates are at the order of
10$^{-7}$\,s$^{-1}$. Direct rotational excitation by $\lambda$ $\sim$ 1\,mm 
photons from the dust is inefficient, because $\tau_{\rm dust}$ $\ll$ 1. For 
vibrational exitation by 4.7$\mu$m photons we obtain with the Einstein 
coefficients $A_{\rm rot}$ = 6.1\,10$^{-6}$ and 3.4\,10$^{-5}$\,s$^{-1}$ for 
the $J$ = 4--3 and 7--6 rotational transitions, $A_{\rm v=1\rightarrow 0}$ 
$\sim$ 20\,s$^{-1}$ (Kirby-Docken \& Liu 1978), and the condition 
$T_{\rm dust}$ $\ga$ 3070\,K/ln\,[$A_{\rm v=1\rightarrow 0}$/$A_{\rm rot}$] 
(see Eq.\,6 of Carroll \& Goldsmith 1981; $\nu_{\rm v=1\rightarrow 0}$ $\sim$ 
6.4\,10$^{13}$\,Hz) a minimum dust temperature of $T_{\rm dust}$ $\sim$ 200\,K. 
This is much larger than $T_{\rm dust}$ = 48\,K, estimated by Hughes et al. 
(1994) and Colbert et al. (1999). The condition for efficient pumping by 
infrared photons is therefore, if at all, only fulfilled for a small part 
of the molecular complex in the central region of M\,82 (cf. McLeod et al. 
1993). Adopting in spite of this $T_{\rm dust}$ $\sim$ 200\,K, the radiative 
pumping rate becomes $B_{\rm v=0\rightarrow 1}$ $u_{\rm v=1-0}$ $\sim$ 
10$^{-15}$\,s$^{-1}$ (or less, if beam dilution plays a role). This is 
negligible in comparison with the collisional pumping rate.

\begin{acknowledgement} 
It is a pleasure to thank the HHT staff, in particular H. Butner, B. Hayward,
D. Muders, F. Patt, and B. Stupak, for their enthusiastic support of the
project and for their flexibility in changing schedules according to variable
weather conditions. For the permission to use the HEB, we also thank the
Harvard-Smithonian Center for Astrophysics (CfA). We acknowledge useful
discussions with S. H{\"u}ttemeister, E. Ros, P. Schilke, C.M. Walmsley, A.
Wei{\ss}, and the useful comments of an anonymous referee. R.Q.M. 
acknowledges support by the 
exchange program between the Chinese Academy of Sciences and the 
Max-Planck-Gesellschaft; C.H. acknowledges support from NATO grant SA.5-2-05
(CRG. 960086) 318/96.
\end{acknowledgement}

\end{document}